\documentclass[mathtimes]{statsoc}
  
\usepackage{amsmath,amssymb,epsf} 

\newcommand{\yy}{{\bf y}} 
\newcommand{\dd}{{\bf d}} 
\newcommand{\YY}{{\bf Y}} 
 
\newcommand{\cB}{{\cal B}} 
\newcommand{\cM}{{\cal M}} 
\newcommand{\cF}{{\cal F}}

\newcommand{\R}{{\mathbb R}}

\newcommand{\cS}{{\cal S}}

\newcommand{\EE}{{\mathbb E}} 
\newcommand{\PP}{{\mathbb P}} 
 
\newcommand{\1}{{\bf 1}}

\newcommand{\mnras}{Monthly Notices of the Royal Astronomical Society} 
\newcommand{\apj}{The Astrophysical Journal}

\title[Cosmic filaments]{A three dimensional object point process for 
detection of cosmic filaments 
\thanks{\emph{Address for correspondence}: Vicent J. Mart{\'\i}nez, 
Observatori Astron\`omic de la Universitat de Val\`encia, 
Edifici d'Instituts d'Investigaci\'o, 
Pol\'{\i}gon La Coma s/n, 46980 Paterna, Val\`encia, Spain.
\textsf{E-mail: martinez@uv.es}}} 
\author[Radu S. Stoica {\it et al.}]{Radu S. Stoica}% 
    \address{Universit\'e Lille 1, Laboratoire Paul Painlev\'e,
59655 Villeneuve d'Ascq Cedex, France.} 
\email{radu.stoica@math.univ-lille1.fr} 
\author{Vicent J. Mart\'{\i}nez} 
    \address{Observatori Astron\`omic de la Universitat de 
Val\`encia, Edifici d'Instituts d'Investigaci\'o, 
Pol\'{\i}gon La Coma s/n, 46980 Paterna, Val\`encia, Spain.}
   \email{martinez@uv.es} 
\author[Radu S. Stoica {\it et al.}]{Enn Saar} 
    \address{Tartu Observatoorium, T\~oravere, 61602 Estonia} 
\email{saar@aai.ee} 
\begin{document} 
 
\begin{abstract} 
We propose to apply an object point process to automatically 
delineate filaments of the large-scale structure in redshift 
catalogues. We illustrate the feasibility of the idea on an example 
of the recent 2dF Galaxy Redshift Survey, describe the procedure, 
and characterise the results. 
\end{abstract} 
 
\section{Introduction} 
 
%%The distribution of galaxies in space is far from random, In fact, 
The large-scale structure of the Universe traced by the 
three-dimensional distribution of galaxies shows intriguing 
patterns: filaments and sheet-like structures connecting in huge 
clusters surround nearly empty regions, the so-called voids. Surveys 
of galaxies are created by measuring for each galaxy in addition to 
its angular position on the sky its distance, estimated from their 
recession velocities, within the framework of a cosmological model. 
However, the measured recession velocities are not only due to the 
Hubble expansion, they appear contaminated by the line-of-sight 
contribution of the dynamical velocity of a galaxy, due to the local 
gravitational effects, so the distances of galaxies are in error. 
Galaxy surveys based on recession velocities are called 'redshift 
space' maps, and although they are distorted versions of the 
three-dimensional distribution of galaxies, distance errors are not 
as serious as to change the overall picture of the large-scale 
structure. 
 
\begin{figure} 
\centering 
\resizebox{0.8\textwidth}{!}{\includegraphics*{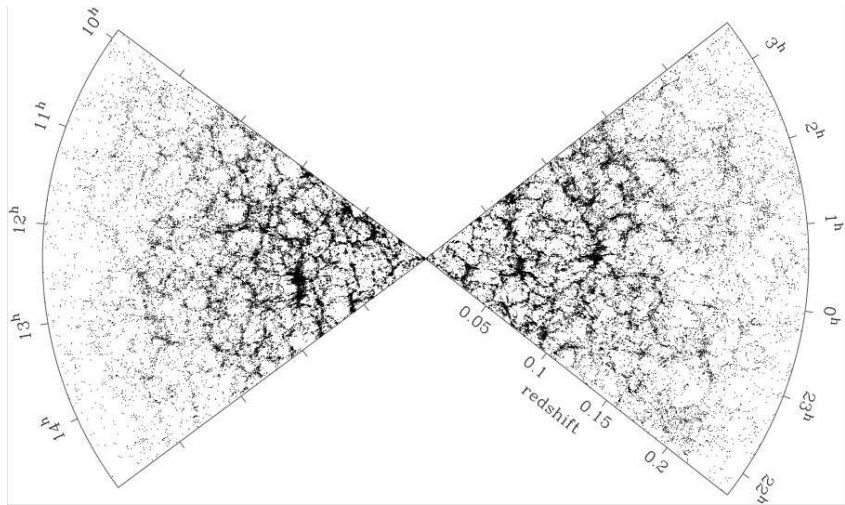}}\\ 
\caption{Galaxy map for two 2dFGRS slices. The observers (we) are 
situated at the centre of the figure. Both slices are thin, with the 
thickness of $2.6^\circ$. The distances are given in redshifts $z$; 
approximately, the physical distance $D\approx 
3000\,h^{-1}\,z\,\mbox{Mpc}$. The numbers along the arcs show the 
right ascension (in hours). The filamentary network of galaxies is 
clearly seen; the disappearance of structure with depth (towards the 
sides of the figure) is caused by luminosity selection. 
\label{fig:2dfgrs} } 
\end{figure}

An overview of such galaxy maps is given in \cite{martsaar02}. As an 
example, we present here a map from a recently completed 2dF Galaxy 
Redshift Survey (2dFGRS, \cite{2dFGRS}). This survey measured the 
redshifts (recession velocities) of galaxies in about 1500 square 
degrees, up to the distances of about 
$700\,h^{-1}\mbox{Mpc}$\footnote{ Distances between galaxies are 
usually measured in megaparsecs (Mpc); 
$1\mbox{Mpc}\approx3\cdot10^{24}\mbox{cm}$. The constant $h$ is the 
dimensionless Hubble parameter; the latest determinations give for 
its value $h\approx 0.71$. } (corresponding to a redshift $z=0.2$ 
for the standard cosmological model). The redshifts were measured in 
two different regions of the sky; Fig.~\ref{fig:2dfgrs} shows the 
positions of galaxies in two $2.6^\circ$ thick slices from both 
regions.

The most characteristic feature of all three-dimensional galaxy 
cartographies are the remarkable network of filaments that can be 
appreciated when diagrams like the one shown in 
Fig.~\ref{fig:2dfgrs} are depicted. All filaments are not equal, 
they show different size and contrast, but typically relatively 
empty voids are found between them. Different scales are 
involved in the filamentary patterns, and it is of great importance 
to have algorithms to identify them and to characterise their 
properties. We should mention here that the gradual disappearance of 
structures with distance observed in Fig.~\ref{fig:2dfgrs} is a 
selection effect due to how this surveys are built. They are 
flux-limited samples, and since the apparent luminosity of a galaxy 
is fainter if the galaxy lies further away, only the brightest 
galaxies are seen in the more distant regions of the surveyed 
volume. 
 
Certainly filaments are prominent and visually they dominate the 
galaxy maps, however there are still no standard methods to 
describe the observed filamentary structure. Second-order summary 
statistics like the two-point correlation function, the $K$-function 
or the power spectrum (in Fourier space) do not provide 
morphological information. Minkowski functionals, minimal spanning 
tree (MST), percolation and shapefinders have been introduced for 
this purpose (for a review see \cite{martsaar02}). 
 
The minimal spanning tree was introduced in cosmology by 
\cite{barrow85}. It is a unique graph that connects all points of 
the process without closed loops, but certainly describes mainly the 
local nearest-neighbour distribution, being unable to provide a whole 
characterisation of the global and large-scale properties of the 
filamentary network. 
 
A better method, named {\it skeleton}, has been recently proposed 
(\cite{eriksen04,novikov06}) to describe the possible filamentary structure of 
continuous density fields. The skeleton is determined by segments 
parallel to the gradient of the field, connecting saddle points to 
local maxima. Calculating the skeleton involves interpolation and 
smoothing the point distribution, which introduces an extra 
parameter, the band-width of the kernel function used to estimate 
the density field from the point distribution, typically a Gaussian 
function. In any case, this method has been recently applied to the 
Sloan Digital Sky Survey (\cite{sousbie06}, providing, by means of 
the length of the skeleton, a good discriminant tool for the 
analysis of the filamentary structures. 
 
In a previous paper (\cite{StoiMartMateSaar05}), we proposed to use 
an automated method to trace filaments for realisations of point 
processes, that has been shown to work well for detection of road 
networks in remote sensing situations 
(\cite{LacoDescZeru05,StoiDescLiesZeru02,StoiDescZeru04}). This 
method is based on the Candy model, a marked point process where 
segments serve as marks. The Candy model can be applied to 2-D 
filaments, and we tested it on simulated galaxy distributions. The 
filaments we found delineated well the filaments detected by eye. 
 
Many methods used to automatically detect filaments have been 
developed so far for two-dimensional maps. This is a natural 
approach for studying the cosmic microwave sky background 
(\cite{eriksen04}), which is two-dimensional, but galaxy maps are 
three-dimensional. The previous filament studies (e.g., 
\cite{bhavsar03, bharadwaj04,pandey05}) have also used 
two-dimensional galaxy maps, projections for thin slices. The main 
reason for that is that most of the past large-scale galaxy maps 
were observed for relatively thin spatial slices; also, in 
projection filaments seem more prominent. But, of course, both 
slicing of filaments and projecting the distribution onto a plane 
distorts the geometry and the properties of the filamentary network. 
 
The study of the three-dimensional filamentary network has just 
begun. Three-dimensio\-nal filaments have been extracted from galaxy 
distribution as a result of special observational projects 
(\cite{pimbblet04a}), or by searching for filaments in the 2dFGRS 
catalogue (\cite{pimbblet04b}). These filaments have been searched 
for between galaxy clusters, determining the density distribution 
and deciding if it is filamentary, individually for every filament. 
Similar studies have been carried out for N-body simulations 
(\cite{colberg05}). A review of these and previous studies of 
large-scale filaments is given in (\cite{pimbblet05}). 
 
No automated methods to trace filaments in the three-dimensional 
galaxy maps have been proposed so far. Based on our previous 
experience with the Candy process, we generalised the approach 
for three dimensions. As the interactions between the structure 
elements are more complex in three dimensions, 
we had to define a more complex model, 
the Bisous model (\cite{StoiGregMate05}). This model gives a general 
framework for the construction of complex patterns made of simple 
interacting objects. In our case, it can be seen as a generalisation 
of the Candy model. We will describe the 
Bisous model below and will apply it to the samples chosen from the 
real three-dimensional galaxy distribution, the 2dFGRS catalogue.

\section{Object point processes: definitions and manipulation tools} 
\subsection{Definitions} 
Let $(K,\cB,\nu)$ be a measure space, where $K$ is a compact subset 
of $\R^3$ with strictly positive Lebesgue measure $0 < \nu(K) < \infty$ 
and $\cB$ the associated Borel $\sigma-$algebra of subsets of K. If, 
to points in $K$, shape descriptors or marks are attached, objects 
are formed. Let $(M,\cM,\nu_{M})$ be the probability measure space of 
these marks. 
 
The considered configuration space is $\Omega = \cup_{n = 
0}^{\infty} \Xi_{n}$, with $\Xi_{n}$ the set of all unordered 
configurations 
$\yy=\{(k_{1},m_{1}),(k_{2},m_{2}),\ldots,(k_{n},m_{n})\}$ that 
consist of $n$ not necessarily distinct objects $y_{i}=(k_{i},m_{i}) 
\in K \times M$. $\Xi_{0}$ is the empty configuration. $\Omega$ is 
equipped with the $\sigma-$algebra $\cF$ generated by the mappings 
that count the number of objects in Borel sets $A \subseteq K \times 
M$. 
 
A marked point process with locations of objects in $K$ and marks in $M$ is 
a measurable mapping from some probability space into $(\Omega,\cF)$. 
An object point process is a marked point process with marks representing 
shape descriptors of the objects.  
 
For simplicity, throughout this paper the shape of an 
object is defined by a compact set $s(y_i)=s(k_i,m_i)$ that is a 
subset of $\R^{3}$ of finite volume $\nu(s(y_i))$. The shape of the object 
configuration $\yy$ is defined by the random set $Z(\yy) = 
\cup_{i=1}^{n(\yy)}s(y_i)$.
 
The simplest object point process is the Poisson object point process. This 
process is used as a reference measure. It generates a configuration 
of objects as follows: first the number of objects is chosen 
according to a Poisson law of intensity $\nu(K)$, then the 
locations of objects are distributed uniformly in $K$ and finally the 
corresponding shapes are chosen independently for each object with 
respect to $\nu_{M}$. 
 
The Poisson object point process does not take into account interactions between 
objects. Indeed, more realistic models can be constructed by 
specifying a probability density with respect to the reference measure: 
\begin{equation} 
p(\yy | \theta) = \alpha \exp\left[-U(\yy|\theta)\right] 
\label{opp_probability_density} 
\end{equation} 
with  the normalising constant $\alpha$,  the model 
parameter vector $\theta$ and the energy function of the system $U(\yy|\theta)$. 
There is a lot of freedom for constructing the energy function, 
provided there exists a positive real constant $\Lambda > 0$ such that 
\begin{equation} 
U(\yy|\theta) - U (\yy \cup \{(k,m)|\theta)\} \leq \log\Lambda 
\label{local_stability} 
\end{equation} 
 
The condition (\ref{local_stability}) is known in the literature as 
the local stability property and it implies the integrability with 
respect to the reference measure, of the probability density given 
by~(\ref{opp_probability_density}). Furthermore, the local stability 
property is of major importance in establishing convergence proofs 
for the Monte Carlo dynamics simulating such a model~\cite{Geye99}. 
The quantity $\exp[U(\yy|\theta)-U (\yy \cup \{(k,m)|\theta)]$ is 
usually called the Papangelou or the conditional intensity ratio. 
 
For further details related to the definition and the properties of 
object point processes we recommend the reader the following 
monographs~(\cite{Lies00,MollWaag03,Reis93}). 
 
\subsection{Manipulation tools: sampling and inference} 
Several Markov chain Monte Carlo Markov techniques are available to 
simulate object point processes~(\cite{GeyeMoll94,Geye99,Gree95,KendMoll00,Lies00,LiesStoi06,Pres77}). 
 
In this paper, we need to sample from the joint law $p(\yy,\theta)$. 
This is done using an iterative Monte Carlo algorithm. An iteration 
of the algorithm consists of two steps. First, a parameter value is 
chosen with respect to $p(\theta)$. Then, conditionally on $\theta$, 
the pattern is sampled from $p(\yy|\theta)$ using a 
Metropolis-Hastings algorithm~(\cite{LiesStoi03,StoiDescZeru04,StoiGregMate05,StoiMartMateSaar05}).  
 
For the problem on hand, $\dd$, the data to be analysed, consist of 
points (galaxies) spread in a finite volume $K$. We want to detect 
the filamentary pattern ``hidden" in these data. Two hypotheses are 
assumed. First, this rather complex pattern is supposed to be formed by 
simple interacting objects. Second, the filamentary pattern is 
considered to be a realisation of an object point process. The energy 
function of such a process can be written as follows: 
\begin{equation} 
U(\yy|\theta)=U_{\dd}(\yy|\theta)+U_{i}(\yy|\theta) 
\label{total_energy} 
\end{equation} 
where $U_{\dd}(\yy|\theta)$ is the data energy and 
$U_{i}(\yy|\theta)$ the interaction energy. 
 
The estimator of the filamentary structure in a field of galaxies 
together with the parameter estimates are given by the configuration 
of objects minimising the total energy function of the system 
\begin{eqnarray} 
(\widehat{\yy},\widehat{\theta}) & = &\arg\max_{\Omega \times \Psi} 
p(\yy,\theta)  = \arg\max_{\Omega \times \Psi}p(\yy|\theta)p(\theta) \nonumber\\ 
& = & \arg\min_{\Omega \times \Psi}\{ 
U_{\dd}(\yy|\theta)+U_{i}(\yy|\theta) - \log p(\theta)\} 
\label{estimator_pattern} 
\end{eqnarray} 
with $\Psi$ the model parameters space. 
 
The minimisation of the energy function can be performed by means of 
a simulated annealing algorithm~(\cite{Lies94,StoiGregMate05}). This 
method iteratively samples from $p(\yy,\theta)^{1/T}$ while slowly 
decreasing the temperature parameter $T$. When the system is frozen 
{\it i.e.} $T \rightarrow 0$, the simulated annealing samples 
uniformly on the sub-space of configurations minimising the energy 
function~(\ref{total_energy}). 
 
In this case, the solution obtained is not unique. Hence, it is 
legitimate to ask if an element of the pattern really belongs to the 
pattern, or if its presence is due to random 
effects~(\cite{StoiGregMate05}). 
 
For compact regions $\cS \subseteq \R^{3}$ of finite volume $0 \leq 
\nu(\cS) < \infty$, we write the probability that an object from the 
pattern $\yy$ covers $\cS$, as follows: 
\begin{eqnarray} 
\PP\left(\1\left\{ \sum_{i=1}^{n(\YY)} \1 \{\cS \subseteq 
s(Y_i)\} > 0 \right\}\right) 
%& = & \nonumber\\ 
%& = &\int_{\Omega \times \Psi} \1\left\{\sum_{i=1}^{n(\YY)} \1 \{\cS 
%\subseteq s(Y_i)\}\right\} d\PP(\YY,\Theta) 
%\label{cover_probability} 
%\\ 
& = &\EE_{(\YY,\Theta)}\1\left\{\sum_{i=1}^{n(\YY)} \1 \{\cS 
\subseteq s(Y_i)\} >0 \right\} \label{cover_probability} 
%\nonumber 
\end{eqnarray} 
 
The cover probability given by (\ref{cover_probability}) can be 
approximated by its Monte Carlo counterpart: 
\begin{equation} 
\widehat{\PP}\left(\1\left\{ \sum_{i=1}^{n(\YY)} \1 \{\cS \subseteq 
s(Y_i)\} > 0 \right\}\right)=  
\frac{1}{U}\sum_{u=1}^{U} 
\1\left\{\sum_{i=1}^{n(\YY_{u})} \1 \{\cS \subseteq 
s(Y_{ui}) >0 \}\right\} 
\label{mcmc_cover_probability} 
\end{equation} 
where $\YY_{1},...,\YY_{U}$ are obtained sampling from $p(\yy,\theta)$. For the 
pattern $\YY_{u}$, the set with the corresponding objects is 
$\{Y_{ui},i=1,\ldots,n(\YY(u))\}$. 
 
%The mark behavior in the region $\cS$ can be analyzed by computing 
%the quantity: 
%\begin{equation} 
%m_{\cS} = \frac 
%{\EE_{(Y,\Theta)} \sum_{v=1}^{n(\YY)}m_{v}\1\{\cS \subseteq s(Y_v)\}} 
%{\EE_{(Y,\Theta)} \sum_{v=1}^{n(\YY)} 1\{\cS \subseteq s(Y_v)\}} 
%\label{mark_behaviour} 
%\end{equation} 
%with its the corresponding Monte Carlo approximation given by: 
%\begin{equation*} 
%\widehat{m}_{\cS} = 
%\frac{\sum_{u=1}^{U}\sum_{v=1}^{n(\YY_u)}m_{v}\1\{\cS \subseteq 
%s(Y_{uv})\}} 
%{\sum_{u=1}^{U}\sum_{v=1}^{n(\YY_u)}\1\{\cS \subseteq s(Y_{uv})\}} 
%\end{equation*} 
 
%The formulas~(\ref{cover_probability}) and~(\ref{mark_behaviour}) 
%can be seen as tools for studying the average behavior in terms of 
%location and shape of the unknown pattern exhibited by the data . 
 
The formula~(\ref{cover_probability}) can be seen as a tool for 
studying the average behaviour in terms of location and shape of 
the unknown pattern exhibited by the data.

There are different choices for the study of the average behaviour 
of the shape exhibited by the data. For instance, we may compute 
the probability that a given compact region would be covered by 
the union of the objects, or the probability that an object 
intersects the given region. We have discarded these alternatives 
because they would lead to an over-detection of the filamentary 
pattern, similar to morphological dilation, and hence to 
filaments which are too thick.
 
\section{Bisous model applied to filamentary pattern detection} 
Detection of cosmic filaments using marked point processes was first 
performed on two-dimensional simulated 
data by \cite{StoiMartMateSaar05}. The marked point process that was 
used throughout the mentioned work was the Candy model. This marked 
point process is able to simulate and detect two-dimensional linear 
networks. Clearly, for real data analysis, a three-dimensional model 
is needed. 
 
In this paper, we use the Bisous model, which is an object 
point process built to model and analyse general three dimensional 
spatial patterns~(\cite{StoiGregMate05}). The spatial pattern is 
made of simple interacting objects or generating elements. The 
particular property of these objects is that they exhibit a finite 
number $q$ of rigid extremity points $\{e_1,\ldots,e_q\}$. Around 
each object $y$ an attraction region $a(y)$ is defined. This 
attraction region is the disjoint union $a(y) = 
\cup_{u=1}^{q}b(e_u,r_a)$, where $b(e_u,r_a)$ is the ball of fixed 
radius $r_a$ centred in $e_u$. Objects attract each other through 
their attraction region. If they attract each other, then they may 
get connected. Connectivity is a more complex interaction than 
attraction. We will show later in this paper that for instance, 
connectivity takes also into account the orientation between the 
main symmetry axes of the objects. The connected objects form the 
pattern. The object connected to the pattern by means of $s$ 
extremity points is called $s$-connected. Among the interactions 
exhibited by the objects forming the pattern we enumerate 
connectivity, alignment and repulsion. They are to be defined later 
in the next section of the paper. 
 
For a pattern $\yy$ the probability density of the Bisous model can be 
written as follows: 
\begin{equation}
p(\yy|\theta) \propto \left[\prod_{i=1}^{n(\yy)}\beta(y_i)\right] 
\left[\prod_{s=0}^{q}\gamma_{s}^{n_{s}(\yy)} \right] 
\prod_{\kappa \in \Gamma} \gamma_{\kappa}^{n_{\kappa}(\yy)} 
\label{bisous_probability_density} 
\end{equation} 
where $\beta: K \times M \rightarrow \R_{+}$ is a positive 
function and $\gamma_s > 0, \gamma_{\kappa} \in [0,1]$ are the 
interaction model parameters. The function  $\beta(y)$ is 
defined below by $\log\beta(y)=u(y)$, where $u(y)$ is 
given by~\eqref{cylinder_contribution}. $\Gamma$ is a set 
of interactions between objects that are pairwise, local 
(distance based), symmetric and guarantee that the density 
(\ref{bisous_probability_density}) is well defined. 
For each $s$ and $\kappa$ the sufficient statistics $n_{s}(\yy)$ 
and $n_{\kappa}(\yy)$ represent, respectively, the number 
of $s$-connected objects and the number of pairs of objects 
exhibiting the interaction of type $\kappa$ in the configuration $\yy$. 
 
The total energy of the model $U(\yy|\theta)$ is computed taking the 
negative logarithm function of~(\ref{bisous_probability_density}). We 
naturally define its two components, the data energy 
\begin{equation} 
U_{\dd}(\yy|\theta) = - \sum_{i = 1}^{n(\yy)} \log \beta(y_i) 
\label{data_energy} 
\end{equation} 
and the interaction energy 
\begin{equation} 
U_{i}(\yy|\theta) = - \sum_{s=0}^{q}n_{s}(\yy)\log\gamma_{s} - 
\sum_{\kappa \in \Gamma} n_{\kappa}(\yy) \log \gamma_{\kappa}. 
\label{interaction_energy} 
\end{equation} 
 
In the following, we will present the generating element together with its 
corresponding interactions that define~(\ref{interaction_energy}), 
and will define the 
data energy~(\ref{data_energy}), which is 
related to the location of the filamentary pattern among the 
galaxies. 
 
\subsection{Generating element and its interactions} 
The generating element of the filamentary pattern ``hidden" in the 
galaxy data is a random thin cylinder. Such a cylinder has a random 
centre in $K$ and has a fixed radius $r$ and a height $h$, with 
$r\ll h$. Its random mark is given by $\omega$, the orientation vector 
of the cylinder. Its corresponding parameters 
$\omega=\phi(\eta,\tau)$ are uniformly distributed on $M=[0,2\pi) 
\times[0,1]$ such that 
\begin{equation*} 
\omega=(\sqrt{1-\tau^2}\cos(\eta),\sqrt{1-\tau^2}\sin(\eta),\tau). 
\end{equation*} 
 
\begin{figure}[!htbp] 
\centering
\resizebox{.6\textwidth}{!}{\includegraphics*{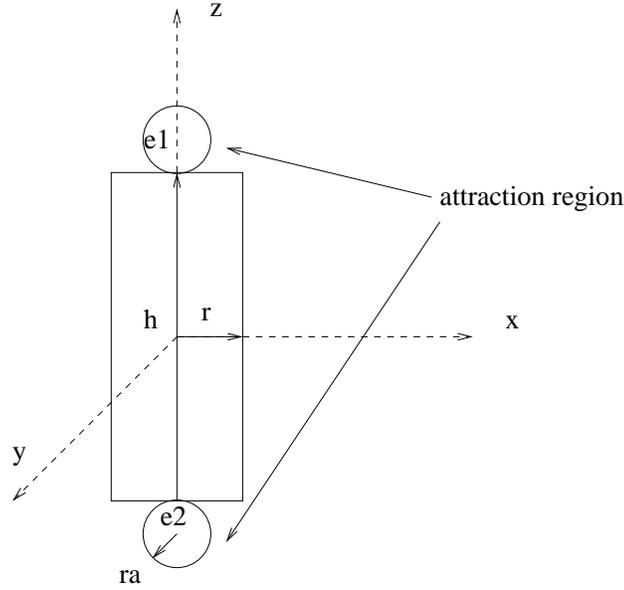}}
\caption{Generating element for the filamentary pattern.} 
\label{cylinder} 
\end{figure}

A cylinder $(k,\omega)$ has $q=2$ extremity points. In 
Figure~\ref{cylinder} the cylinder is centred in the origin and 
its symmetry axis is parallel to $Oz$. The coordinates of the 
extremity points are 
\begin{equation*} 
e_{u} = (0,0,(-1)^{u+1}(\frac{h}{2}+r_a)), \quad u \in \{1,2\} 
\end{equation*} 
and the orientation vector is $\omega=(0,0,1)$. A randomly located 
and oriented cylinder is obtained by the means of a translation and two 
rotations~(\cite{HearBake94}). 
 
Two cylinders $y_{1}=(k_1,\omega_1)$ and $y_{2}=(k_2,\omega_2)$ 
attract each other, and we write $y_1 \sim_{a} y_2$ if the set 
\begin{equation*} 
\{(u,v) : 1 \leq u,v \leq 2, u \neq v, d(e_{u}(y_1),e_{v}(y_2)) \leq 
r_a\} 
\end{equation*} 
contains only one element. 
 
The same cylinders reject each other, and we write $y_{1} \sim_{r} 
y_{2}$ if $d(k_{1},k_{2}) < h$. They are aligned, and we write $y_{1} 
\sim_{\parallel} y_{2}$ if 
\begin{equation*} 
\omega_{1} \cdot \omega_{2} \geq 1 - \tau_{\parallel} 
\end{equation*} 
where $\tau_{\parallel} \in (0,1)$ is a predefined curvature 
parameter and $\cdot$ designs the scalar product of the two 
orientation vectors. If 
\begin{equation*} 
|\omega_{1} \cdot \omega_{2}| \leq \tau_{\perp} 
\end{equation*} 
where $\tau_{\perp} \in (0,1)$ is a predefined crossing curvature, 
they are said to be orthogonal, and we 
write 
$y_{1} \sim_{\perp} y_{2}$. 
 
Two cylinders $y_1$ and $y_2$ are connected and we write $y_1 
\sim_c y_2$ if the following conditions are simultaneously 
fulfilled: 
\begin{equation*} 
\begin{array}{l} 
y_1 \sim_{a} y_2\\ 
y_1 \not\sim_{r} y_2\\ 
y_1 \sim_{\parallel} y_2\\ 
\end{array} 
\end{equation*} 
 
If the following conditions are simultaneously verified: 
\begin{equation*} 
\begin{array}{l} 
y_1 \sim_{r} y_2\\ 
y_1 \not\sim_{\perp} y_2\\ 
\end{array} 
\end{equation*} 
the cylinders exhibit a hard repulsion, and we write $y_1 \sim_{h} 
y_2$. 
 
The filaments forming the pattern are made of cylinders that attract 
each other and they are well aligned. These filaments may cross. At 
the  same time configurations with overlapping cylinders are not 
desirable since they may outline clusters of galaxies instead of 
filaments.  
 
\begin{figure}[!htbp] 
\centering
\resizebox{.6\textwidth}{!}{\includegraphics*{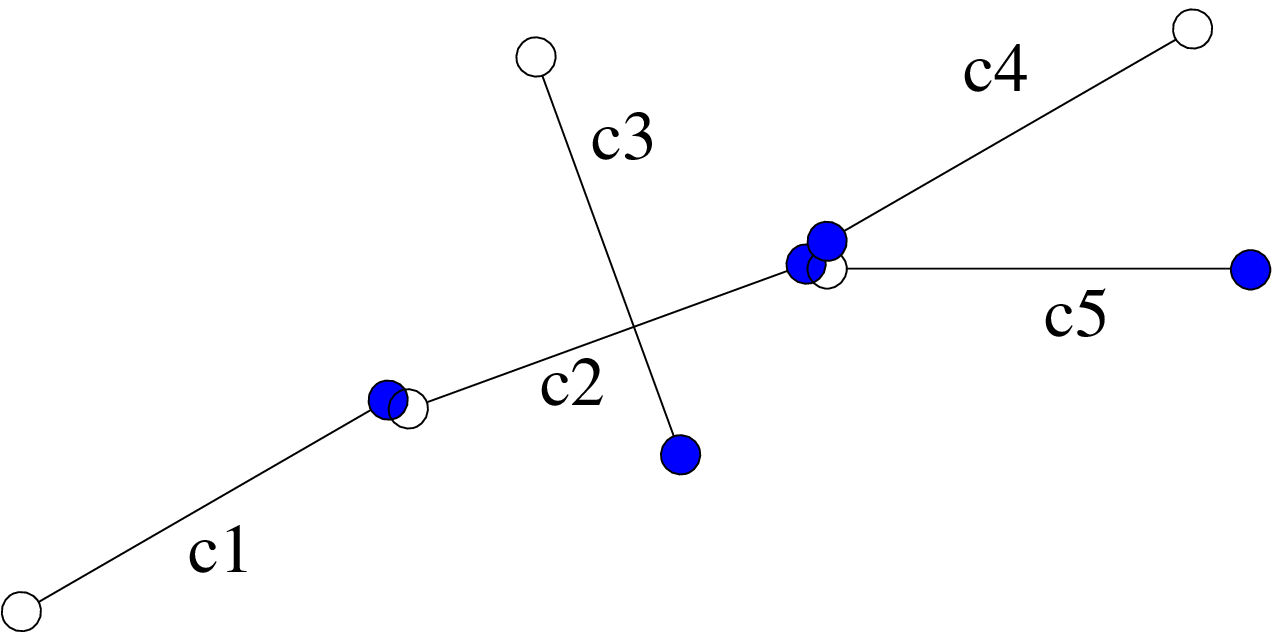}} 
\caption{Two-dimensional representation of a configuration of interacting cylinders.} 
\label{config_cylinder} 
\end{figure}

In Figure~\ref{config_cylinder} a cylinder configuration is represented 
in two dimensions. According to the previous definitions, we observe 
that $c_1 \sim_c c_2$ since $c_1 \sim_a c_2$, $c_1 \not\sim_r c_2$, 
and $c_1 \sim_\parallel c_2$. Thus, the cylinders $c_1$ and $c_2$ are 
$s$-connected. In order to form filaments, this kind of interaction 
is encouraged by the model. The cylinders $c_2$ and $c_3$ reject each other, 
but they are rather orthogonal, so their position is not penalised by 
the model. The cylinders $c_2$ and $c_4$ are not $s$-connected since 
$c_2 \not\sim_a c_4$, and their position is not encouraged nor penalised +
by the model. The cylinders $c_4$ and $c_5$ reject each other, $c_4 \sim_r c_5$,
and simultaneously they tend to overlap, $c_4 \not\sim_{\perp} c_5$. 
So, this pair of cylinders exhibits a hard repulsion interaction and this 
configuration is strongly penalised  by the model.  
 
With respect to these considerations, all the ingredients 
needed for the natural construction of the interaction 
energy~(\ref{interaction_energy}) are defined. The $s$-connectivity 
is constructed using the relation $\sim_c$. The $\Gamma$ set contains 
here only one element given by $\sim_h$. The model parameters for the 
interaction energy are $\gamma_0,\gamma_1,\gamma_2 > 0$ and $\gamma_h 
\in [0,1]$. The definition of the interactions and the parameter ranges 
ensure the object point process induced by the interaction energy to 
be well defined, locally stable and Markov in the sense of 
Ripley-Kelly~(\cite{StoiGregMate05}).

\subsection{The data energy} 
\begin{figure}[!htbp] 
\centering
\resizebox{.6\textwidth}{!}{\includegraphics*{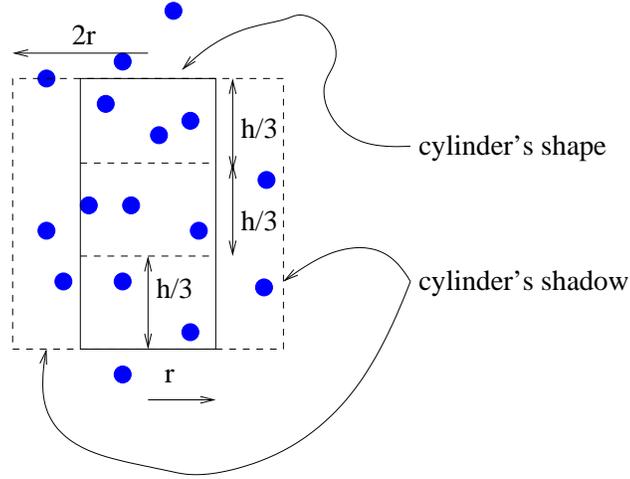}} 
\caption{A two-dimensional view of a cylinder with its shadow in a 
pattern of galaxies.} 
\label{data_cylinder} 
\end{figure}

The data energy is related to the position of the filamentary pattern 
in the galaxy field. To each cylinder $y$ an extra cylinder is 
attached, so that it has exactly the same parameters as $y$, except 
for the radius which equals $2r$.  Let $\tilde{s}(y)$ be the shadow 
of $s(y)$ obtained by the substraction of the initial cylinder from 
the extra cylinder, as indicated in Figure~\ref{data_cylinder}. The 
cylinder $y$ is divided along its main symmetry axis, in three equal 
volumes, and we denote by $s_{1}(y)$, $s_{2}(y)$ and $s_{3}(y)$ their 
corresponding geometrical shapes. 
 
A cylinder may belong to the pattern if several conditions are 
verified. First, the density of galaxies inside $s(y)$ has to be 
higher than the density of galaxies in $\tilde{s}(y)$, and it can be 
written as follows: 
\begin{equation*} 
\1\{\text{``density"}\} = \1 \{ n(\dd \cap s(y))\nu(\tilde{s}(y)) > 
n(\dd \cap \tilde{s}(y))\nu(s(y)) \} 
\label{density_condition} 
\end{equation*} 
where $n(\dd \cap s(y))$ and $n(\dd \cap \tilde{s}(y))$ represent the 
number of galaxies covered by the cylinder and its shadow, 
and $\nu(s(y))$ and $\nu(\tilde{s}(y))$ are the volumes of the cylinder 
and its shadow, respectively. Next, the galaxies covered have to be spread 
inside the whole volume. This is formulated below: 
\begin{equation*} 
\1\{\text{``spread"}\} = \prod_{i=1}^{3}\1\{ n(\dd \cap s_{i}(y)) > 1 
\} 
\label{spread_condition} 
\end{equation*} 
with $n(\dd \cap s_{i}(y))$ the number of galaxies belonging to 
$s_{i}(y)$. Under these assumptions, $u(y)$ the energy contribution 
of a cylinder is defined as follows: 
\begin{equation} 
u(y)=\1\{\text{``density"}\}\1\{\text{``spread"}\}( n(\dd \cap s(y)) 
- n(\dd \cap \tilde{s}(y) + u_{\max}) -u_{\max} 
\label{cylinder_contribution} 
\end{equation} 
where $u_{\max}$ is a positive fixed quantity. The introduction of 
this term gives to a segment that does not meet all the previous conditions,
a very small chance to be a part of the network. This should give more 
complete networks and better mixing properties to the method.  
 
The data energy~(\ref{data_energy}) is obtained by summing the 
energy contributions~(\ref{cylinder_contribution}) for all the 
cylinders in the pattern: 
\begin{equation} 
U_{\dd}(\yy|\theta) = - \sum_{i=1}^{n(\yy)} u(y_i). 
\label{data_term_cylinder} 
\end{equation} 
It can be checked that the data energy term~(\ref{data_term_cylinder}) 
induces a locally stable object point process.

\section{Data} 
 
\subsection{Observational data} 
 
The best available redshift catalogue to study morphology of the 
galaxy distribution at present is the 2dF Galaxy Redshift Survey 
(2dFGRS) \citep{2dFGRS}. The catalogue covers two separate regions 
in the sky, the NGP (North Galactic Cap) strip, and the SGP 
(South Galactic Cap) strip, with total area of about 1500 
square degrees. The nominal (extinction-corrected) magnitude 
limit of the catalogue is $b_j=19.45$; 
reliable redshifts were obtained for 221414 galaxies. 
As seen in Fig.~\ref{fig:2dfgrs}, the effective depth of the 
catalogue is about $z=0.2$ or $D\approx600\,h^{-1}\,$Mpc, 
and the total volume $V\approx3.3\cdot10^7h^{-3}\,$Mpc$^3$. 
 
As all galaxy redshift catalogues, the catalogue is not strictly 
homogeneous. First, due to different observing conditions, the 
magnitude limit (the depth of the catalogue) changes from 
observation to observation, but the changes are known and well 
documented. Second, due to the fact that the fibbers that lead the 
light from the focal plane to the spectrograph have a finite size, 
the redshifts of a few per cent of selected galaxies could not be 
measured. This effect is quantified as redshift completeness. All 
the data and the programs to calculate the magnitude limits and the 
spectroscopic completeness are public and can be found at 
\texttt{http://www.mso.anu.edu.au/2dFGRS/}. 
 
The 2dFGRS catalogue is flux-limited catalogue and therefore the 
density of galaxies decreases with distance. For statistical 
analysis of such surveys, a weighting scheme that compensates for 
the missing galaxies at large distances, has to be used. Usually, 
each galaxy is weighted by the inverse of the selection function 
\citep{martsaar02}. The selection function is determined by the two 
main selection effects described above.  However, such a weighting 
is suitable only for specific statistical problems, as, e.g., the 
calculation of correlation functions. When studying the local 
structure, such a weighting cannot be used; it would only amplify 
the shot noise.

At the cost of discarding many surveyed galaxies, one can 
alternatively use volume-limited samples. In this case, the 
variation in density at different locations depends only on the 
fluctuations of the galaxy distribution itself. We started our 
analysis by using the volume-limited samples prepared by the 2dF 
team for scaling studies (\cite{croton1}) and kindly sent to us by 
Darren Croton. Unfortunately, we found soon that the galaxy density 
in these samples was too low and the shot noise did not allow 
algorithms to find filaments; there are, typically, only of the 
order of ten galaxies per filament is such samples. 
 
\begin{figure} 
\centering 
\resizebox{0.4\textwidth}{!}{\includegraphics*{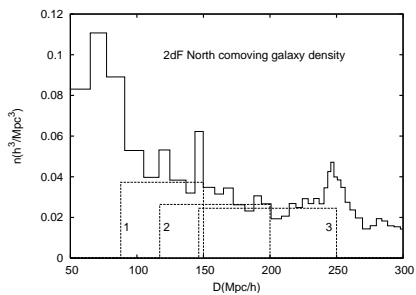}}\\ 
\caption{Galaxy density for the full 2dFGRS NGP sample (in 
comoving coordinates) versus the distance $D$ from the 
observer. As the nearby volume is small, single 
superclusters cause strong spikes in the histogram. The most 
prominent supercluster that creates the spike at $D\approx250\,h^{-1}$Mpc, 
lies just outside of the largest data brick used in this work. 
\label{fig:dens}} 
\end{figure}

Thus we choose another way -- we started from a full galaxy sample, 
and choose from that bricks in a limited distance range, where the 
galaxy density was approximately constant. 
We use bricks in this paper to avoid problems with complex boundaries
of the full 2dfGRS sample.
As the SGP half of the 
galaxy sample has a convex geometry (it is limited by two conical 
sections of different opening angles), the constant density bricks 
that is possible to cut from the SGP have small volumes. Thus we 
used only the NGP data, and chose three bricks of maximum length and 
height, with different depths. General characteristics of these data 
sets are listed in Table~\ref{tab:bricks}. As seen from the 
(comoving) galaxy density histogram in Fig.~\ref{fig:dens}, the 
densities inside the bricks are approximately constant. 
 
\begin{table}
\caption{Galaxy content and geometry for the 2dFGRS bricks (in
$h^{-1}$Mpc). $N_{\mbox{gal}}$ is the number of galaxies in a sample, 
$D_{\min}$
is the distance of a sample from the observer (its nearest point),
$D_{\max}$ is the maximum distance in a sample from
the observer (to the far-away brick corners) and $d$ is the mean
distance between galaxies in a sample. \label{tab:bricks}}\\ 
%\centering 
\begin{tabular}{rrrrrrrr} 
\hline\\ 
sample&$N_{\mbox{gal}}$&$D_{\min}$&depth&width&height&$D_{\max}$&$d$\\[3pt] 
\hline\\ 
NGP150&2499& 87.9&53.1&101.5&12.4&150.0&2.99\\ 
NGP200&4180&117.2&70.9&135.3&16.5&200.0&3.36\\ 
NGP250&7588&146.4&88.6&169.1&20.7&250.0&3.44\\[3pt] 
\hline\\ 
\end{tabular} 
\end{table}

\begin{figure} 
\centering 
\resizebox{0.7\textwidth}{!}{\includegraphics*{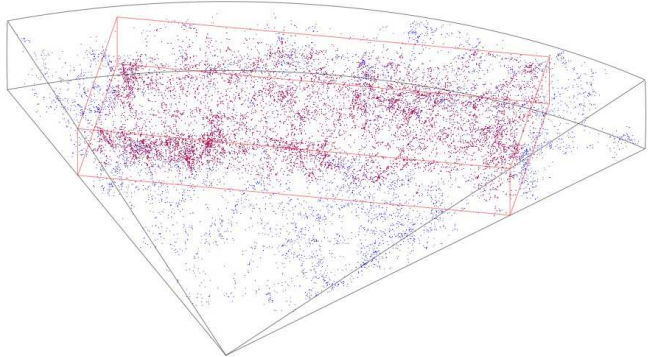}} 
  \caption{One of the cuboidal samples analysed in this paper 
  drawn from the Northern Galactic Cap of the 2dFGRS. 
  The wedge shows the approximate extent of the full sample, its
  real borders are more complex.
\label{fig:2dfpoints}} 
\end{figure}

We show one of these three bricks inside the 2dFGRS NGP sample 
limits and the spatial distribution of galaxies in 
Fig.~\ref{fig:2dfpoints}. 
 
\subsection{Experimental results} 
As described above, we use three data sets, drawn from the galaxy 
distribution in the Northern subsample of the 2dFGRS survey. The 
size of the brick determines the size of $K$. As the data resolution 
is similar (column $d$ in Table~\ref{tab:bricks}, we choose the same 
values for the dimensions of the cylinder for all the data sets, as 
follows: $r=0.5$, $h=6.0$. The radius of the cylinder is close to 
the minimal one can choose, taking into account the data resolution. 
Its height is also close to the shortest possible, as our shadow cylinder has 
to have a cylindrical geometry, too (the ratio of its height to the 
diameter is presently 3:1). We choose the attraction radius as 
$r_{a}=0.5$, giving for the maximum distance between connected 
cylinders the value 1.5, and the pre-defined curvatures are 
$\tau_{\parallel}=\tau_{\perp}=0.15$. This allows for a maximum of 
$\approx30^\circ$ between the direction angles of connected 
cylinders, and limits the orthogonality condition to angles larger 
than $\approx80^\circ$. 
 
The data energy parameter is $u_{\max}=-25$. For the interaction 
energy, the parameter domain is chosen as follows: $\log \gamma_{0} 
\in [-12.5,-7.5]$, $\log \gamma_{1} \in [-5, 0] $ and $\log 
\gamma_{2} \in [0, 5]$. The hard repulsion parameter is 
$\gamma_{h}=0$, so the configurations with cylinders exhibiting such 
interactions are forbidden. The domain of the connection parameters 
was chosen such that $2$-connected cylinders are generally encouraged, 
$1$-connected cylinders are penalised and $0$-connected segments are 
strongly penalised. This choice encourages the cylinders to group in 
filaments in those regions where the data energy is good enough. 
Still, we have no information about the relative strength of those parameters. 
Therefore, we have decided to use for the prior parameter density $p(\theta)$ 
the uniform law over the parameter domain. 
 
%The joint law $p(\yy,\theta)$ is simulated iteratively. An iteration 
%consists of two steps. First, a parameter value is chosen according 
%to $p(\theta)$. Then, conditionally on $\theta$, a new configuration 
%of cylinders is chosen with respect to $p(\yy|\theta)$. The 
%conditional law $p(\yy|\theta)$ was simulated using a tailored to 
%the model Metropolis-Hastings 
%algorithm~(\cite{LiesStoi03,StoiEtAles05}). 
 
A delicate point when building a simulated annealing procedure is 
the choice of an appropriate cooling schedule for the temperature. 
(\cite{StoiGregMate05}) proved that the use of a logarithmic cooling 
schedule guarantees the convergence of the simulated annealing algorithm 
when the parameters of the point process are fixed. Therefore, we have 
chosen the following scheme for the descent of the temperature: 
\begin{equation*} 
T_{n} = \frac{T_{0}}{1+ \log n} 
\end{equation*} 
 
\noindent 
with $T_{0}=1$. 
 
\begin{figure}[!htbp] 
\centering
a)\resizebox{0.6\textwidth}{!}{\includegraphics*{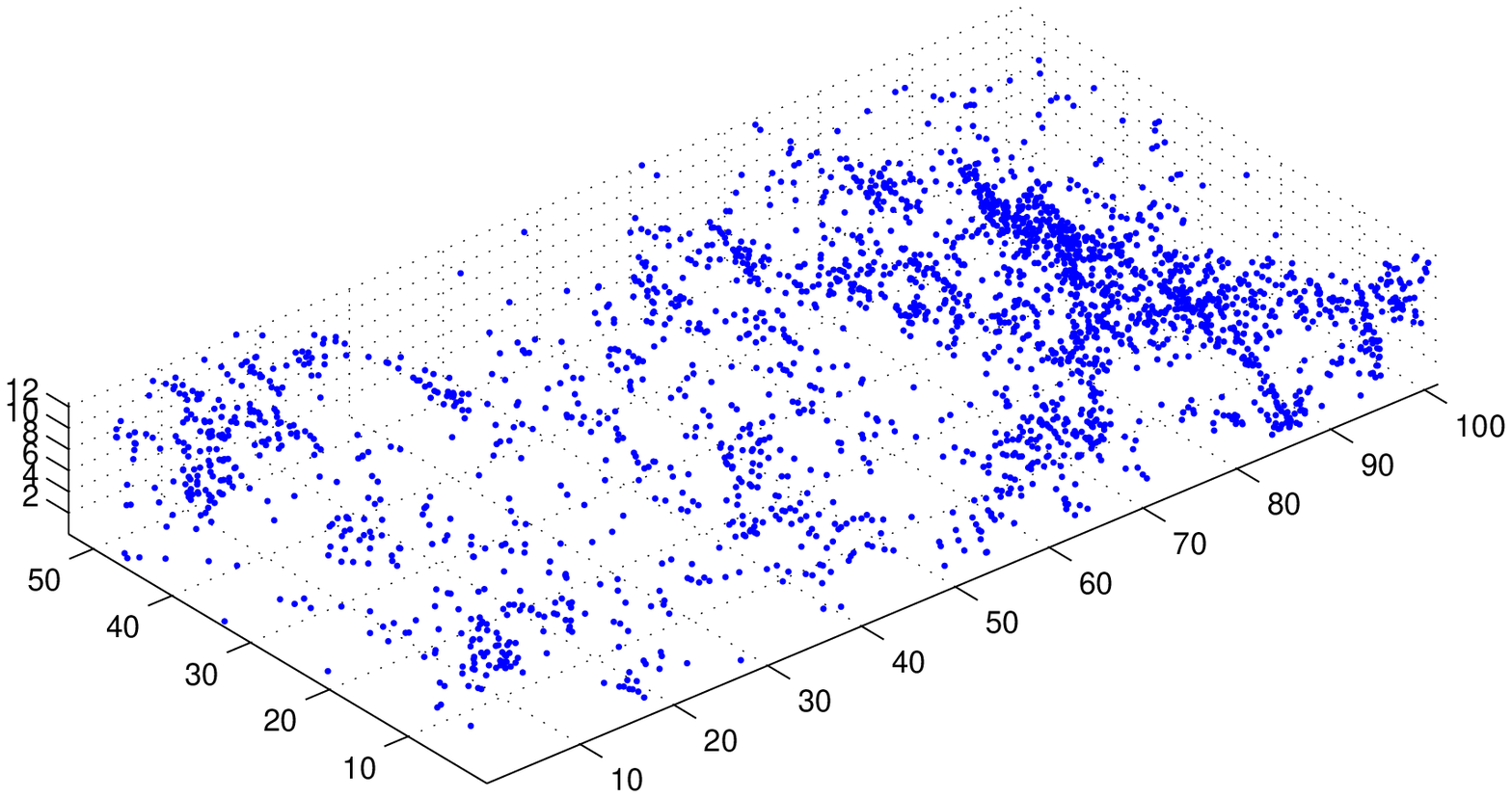}}\\ 
b)\resizebox{0.6\textwidth}{!}{\includegraphics*{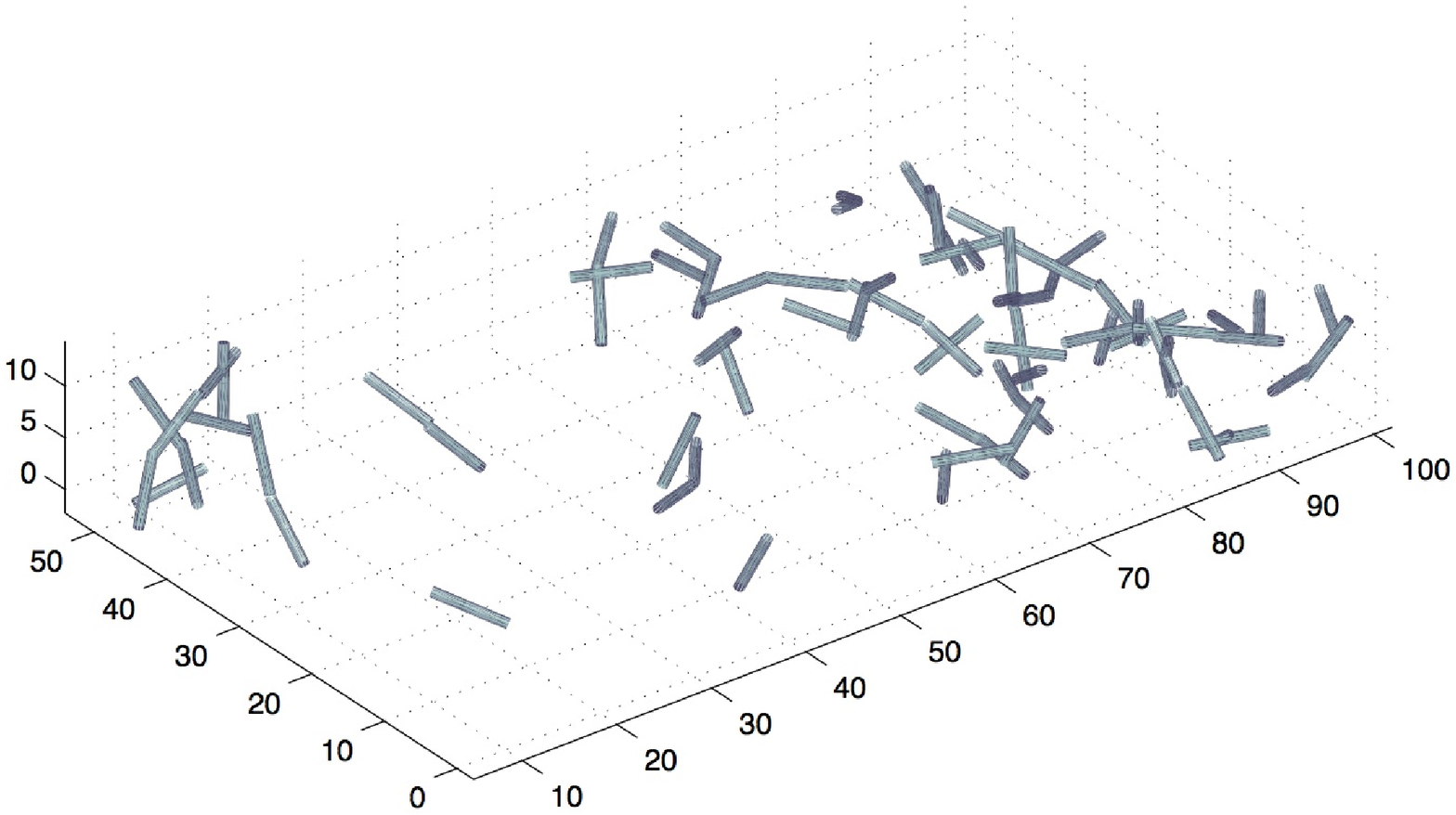}}\\ 
c)\resizebox{0.6\textwidth}{!}{\includegraphics*{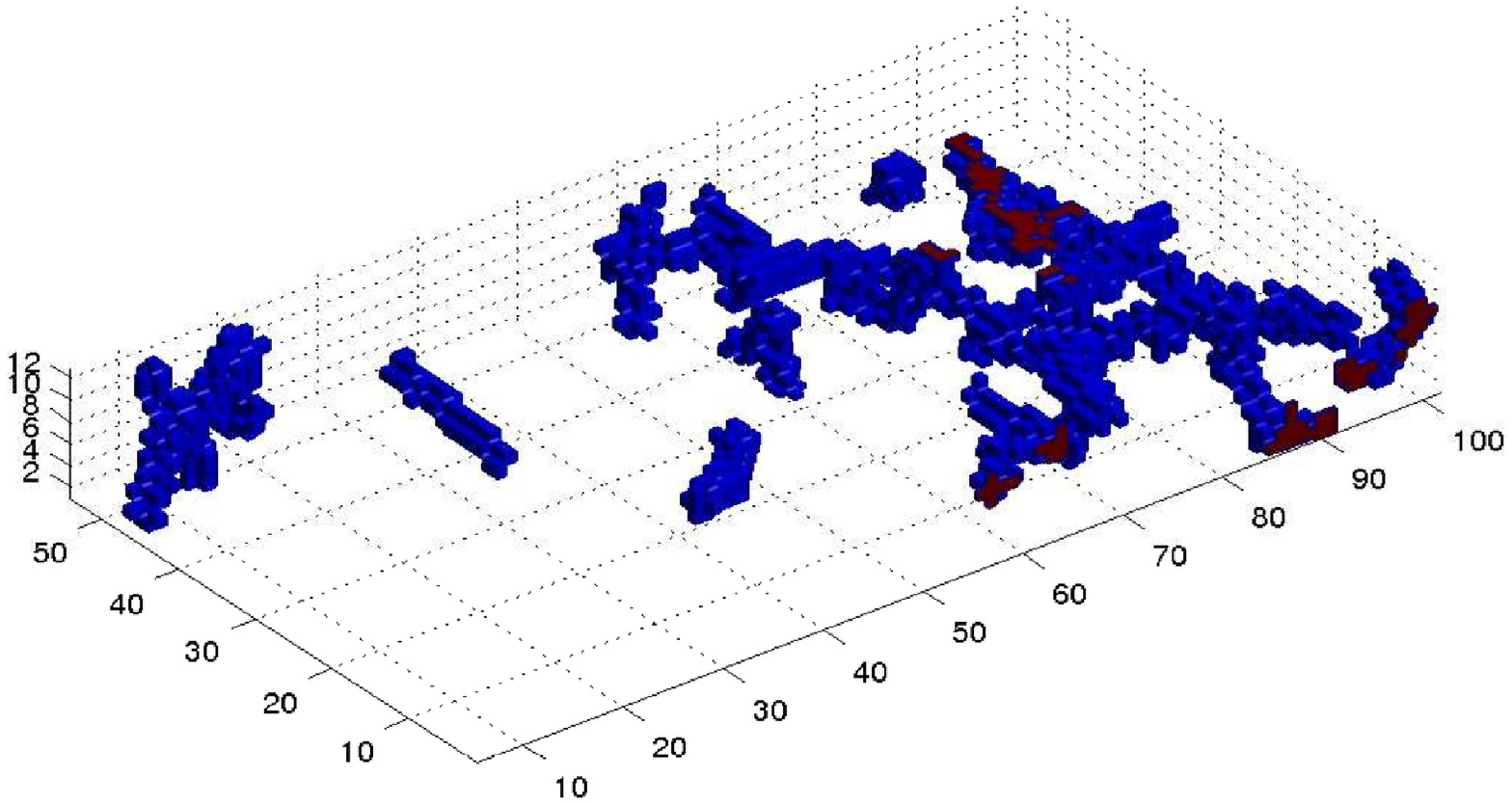}}\\ 
d)\resizebox{0.6\textwidth}{!}{\includegraphics*{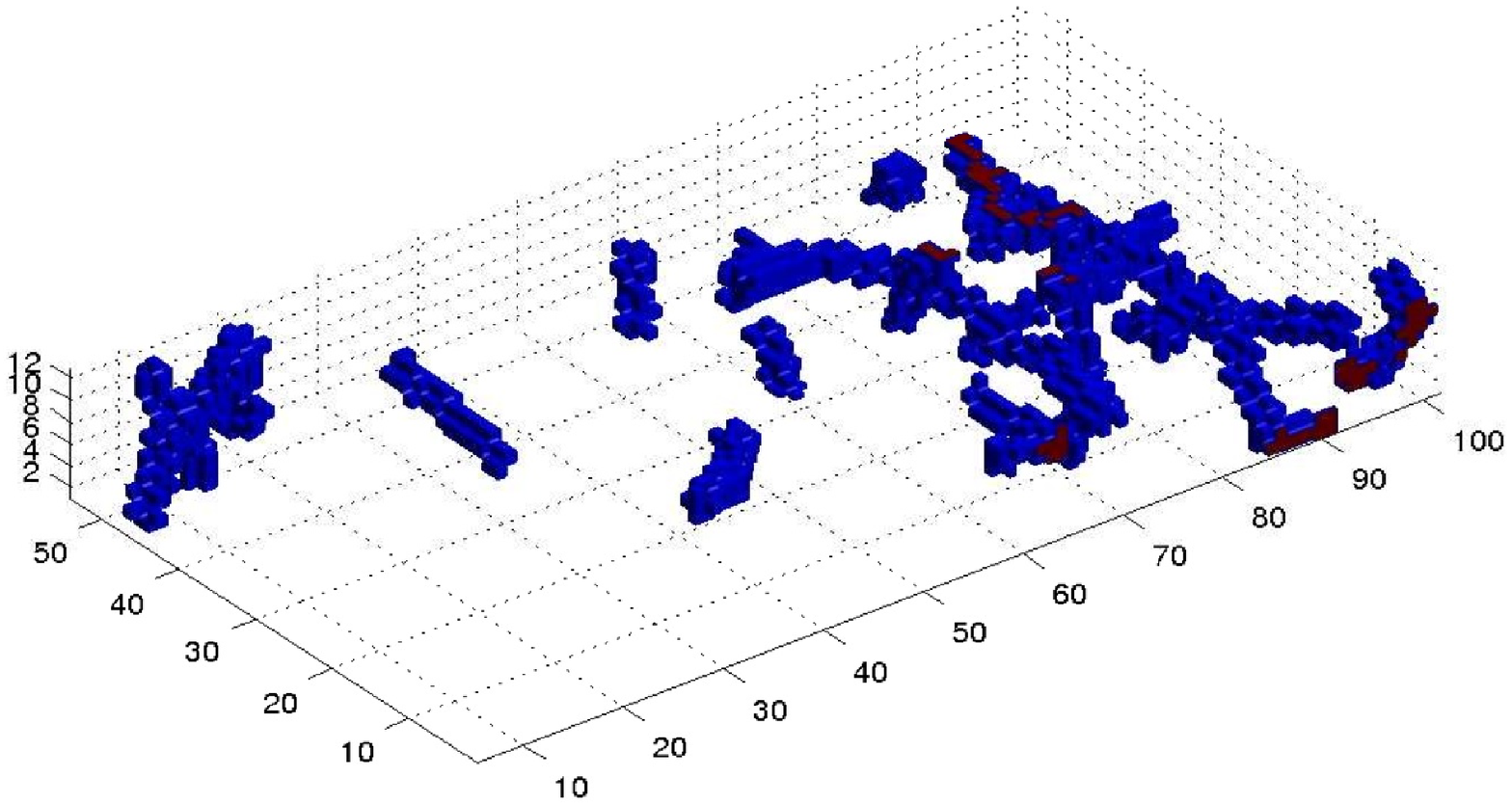}}\\ 
\caption{Data set $NGP150$. a) The data. 
b) Cylinder configuration obtained using the 
simulated annealing algorithm. Cover probabilities thresholded at 
c) $50\%$, 
%d) $75\%$, 
d) $95\%$.} 
\label{res_n150} 
\end{figure}

\begin{figure}[!htbp] 
\centering
a)\resizebox{0.55\textwidth}{!}{\includegraphics*{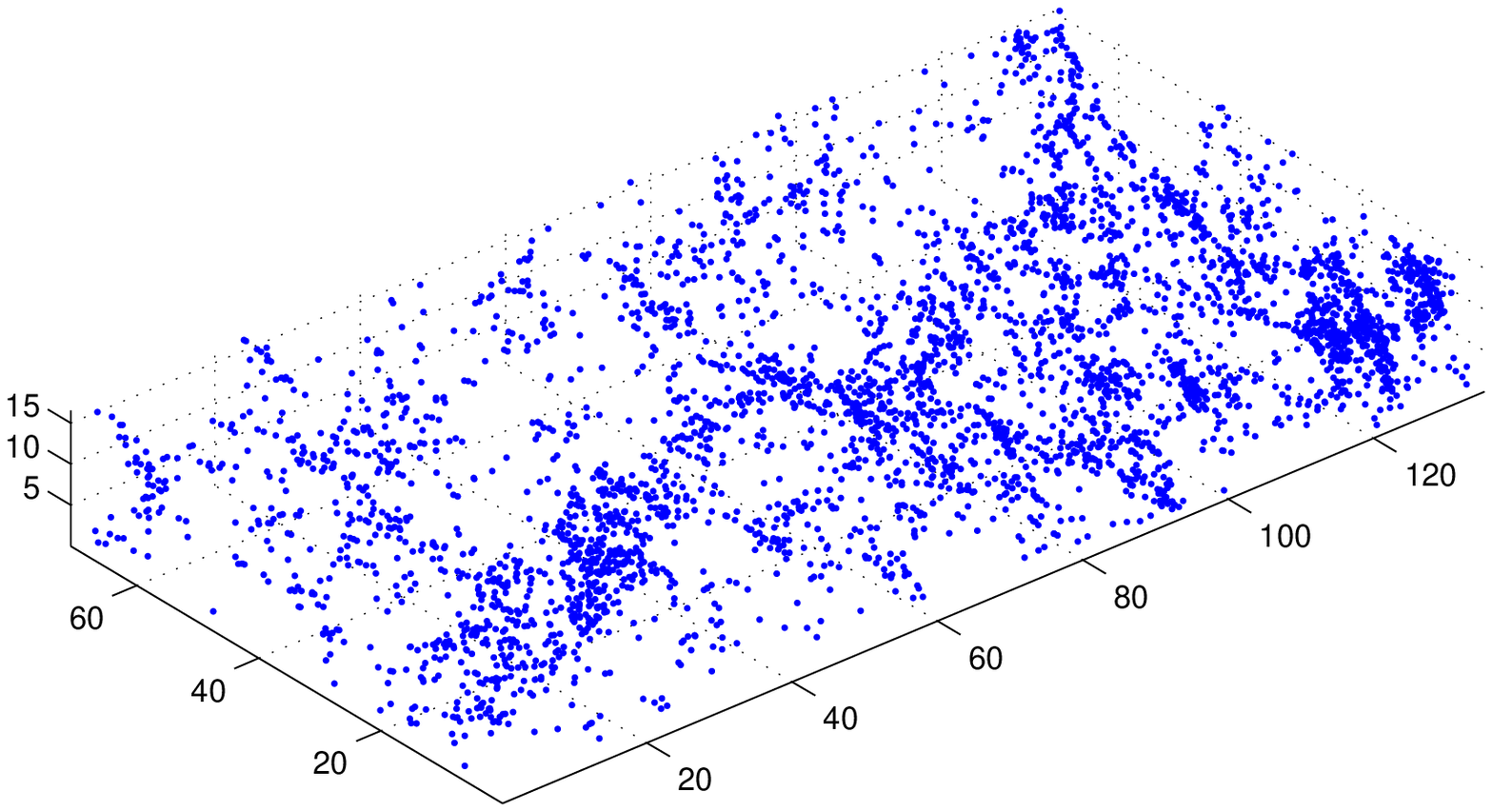}}\\ 
b)\resizebox{0.55\textwidth}{!}{\includegraphics*{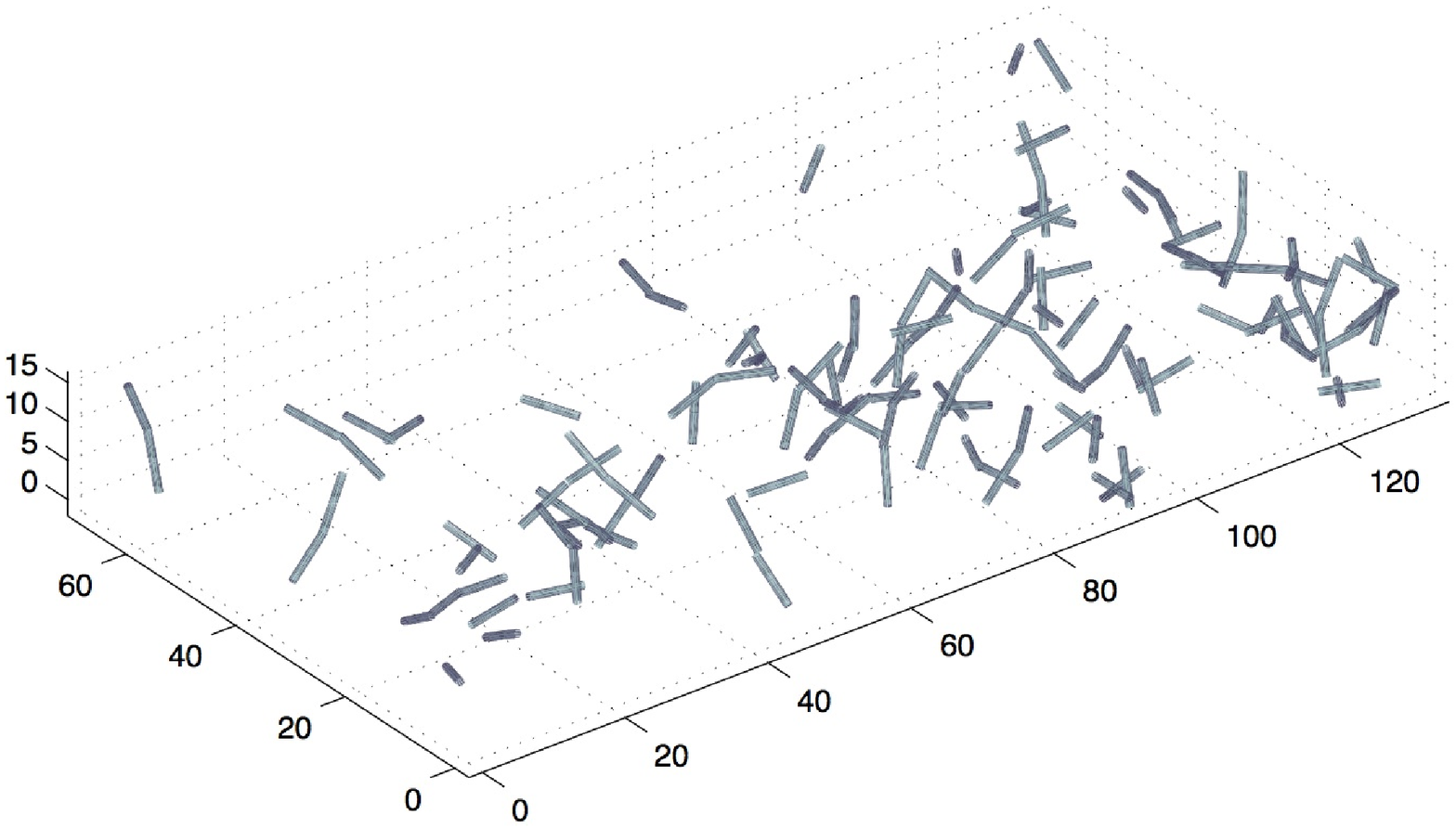}}\\ 
c)\resizebox{0.55\textwidth}{!}{\includegraphics*{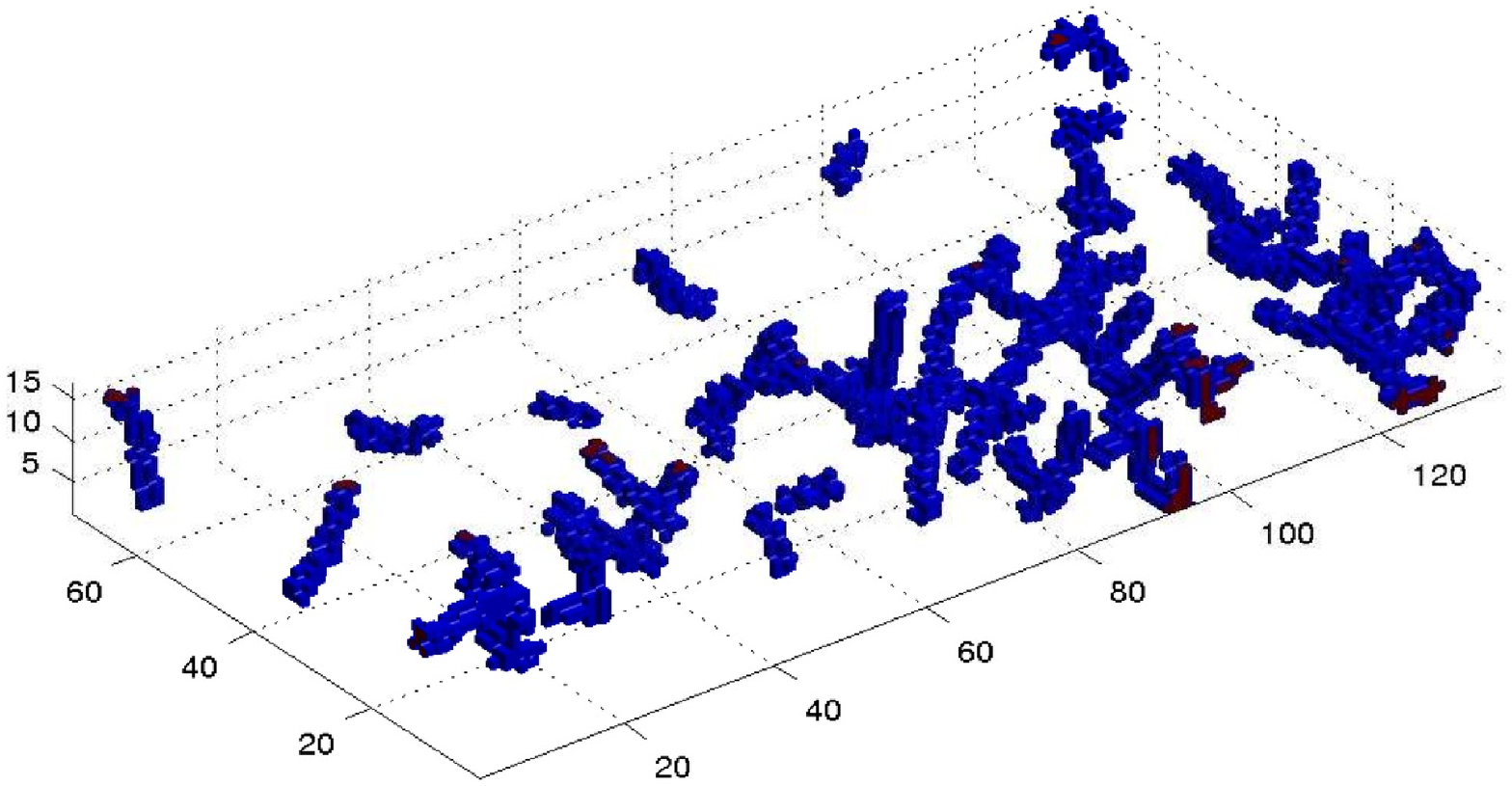}}\\ 
d)\resizebox{0.55\textwidth}{!}{\includegraphics*{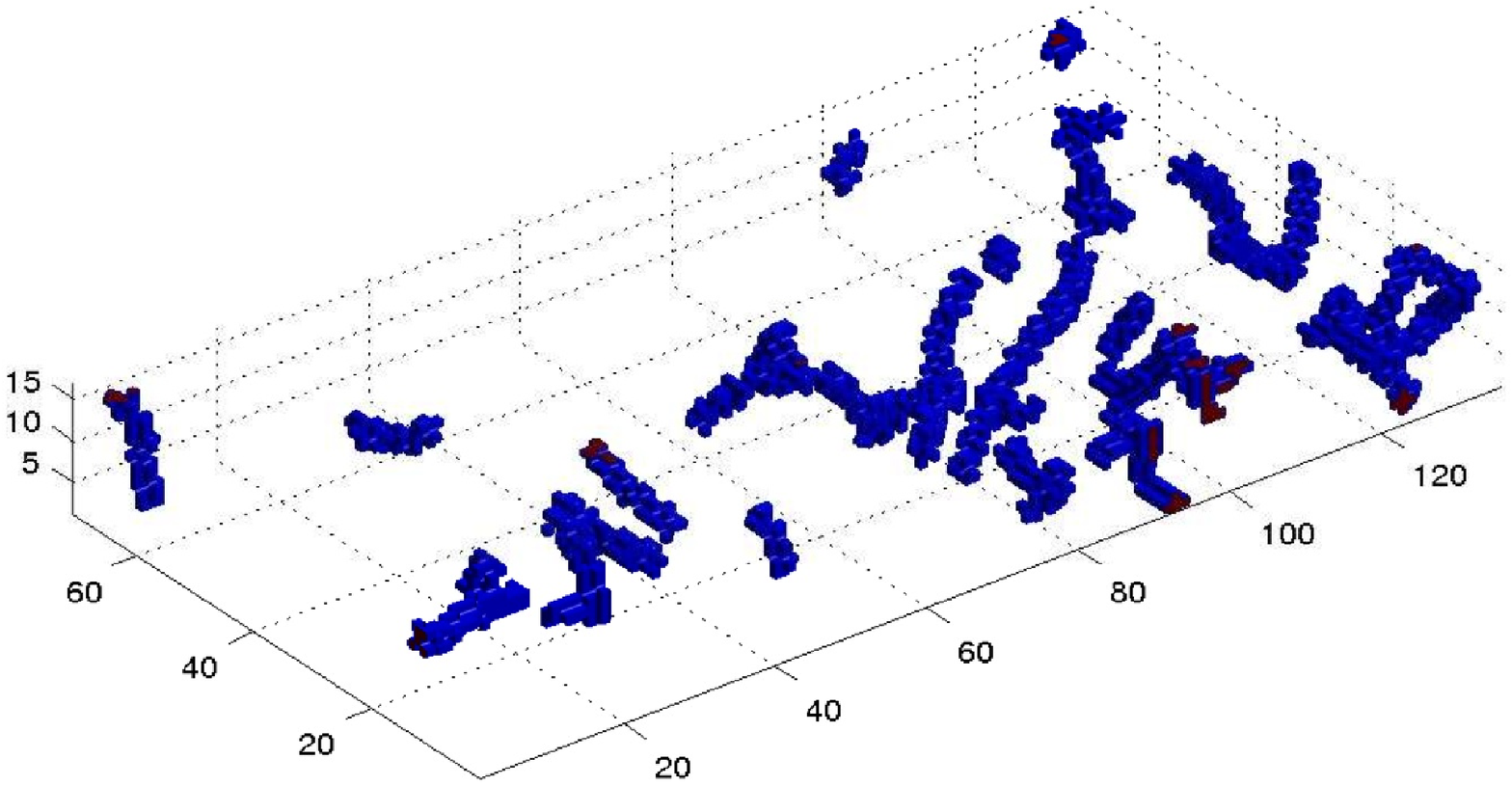}}\\ 
\caption{Data set $NGP200$. a) The data, 
b) Cylinder configuration obtained using the 
simulated annealing algorithm. Cover probabilities thresholded at c) 
$50\%$, %c) $75\%$, 
d) $95\%$.} 
\label{res_n200} 
\end{figure}

\begin{figure}[!htbp] 
\centering
a)\resizebox{0.6\textwidth}{!}{\includegraphics*{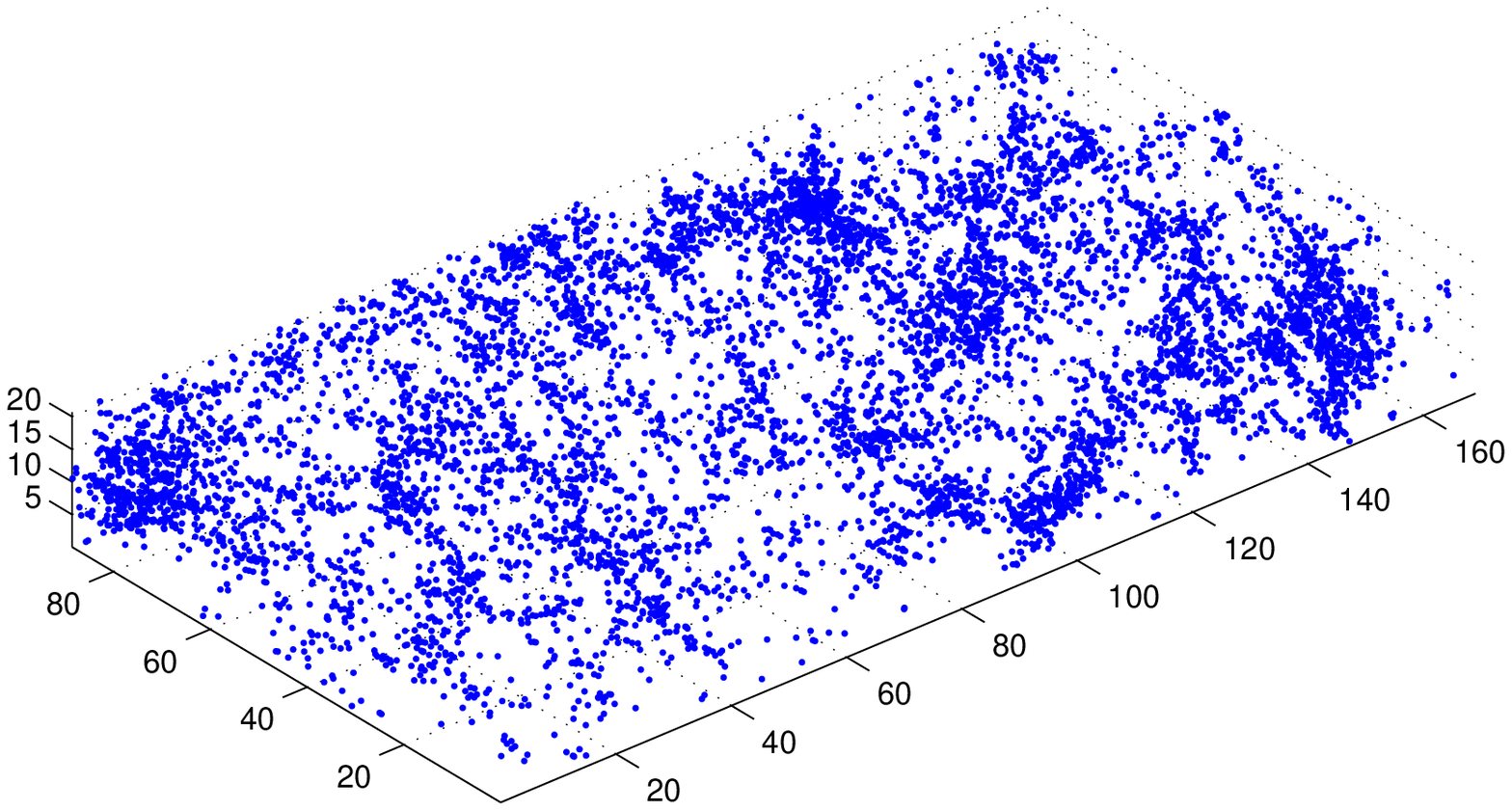}}\\ 
b)\resizebox{0.6\textwidth}{!}{\includegraphics*{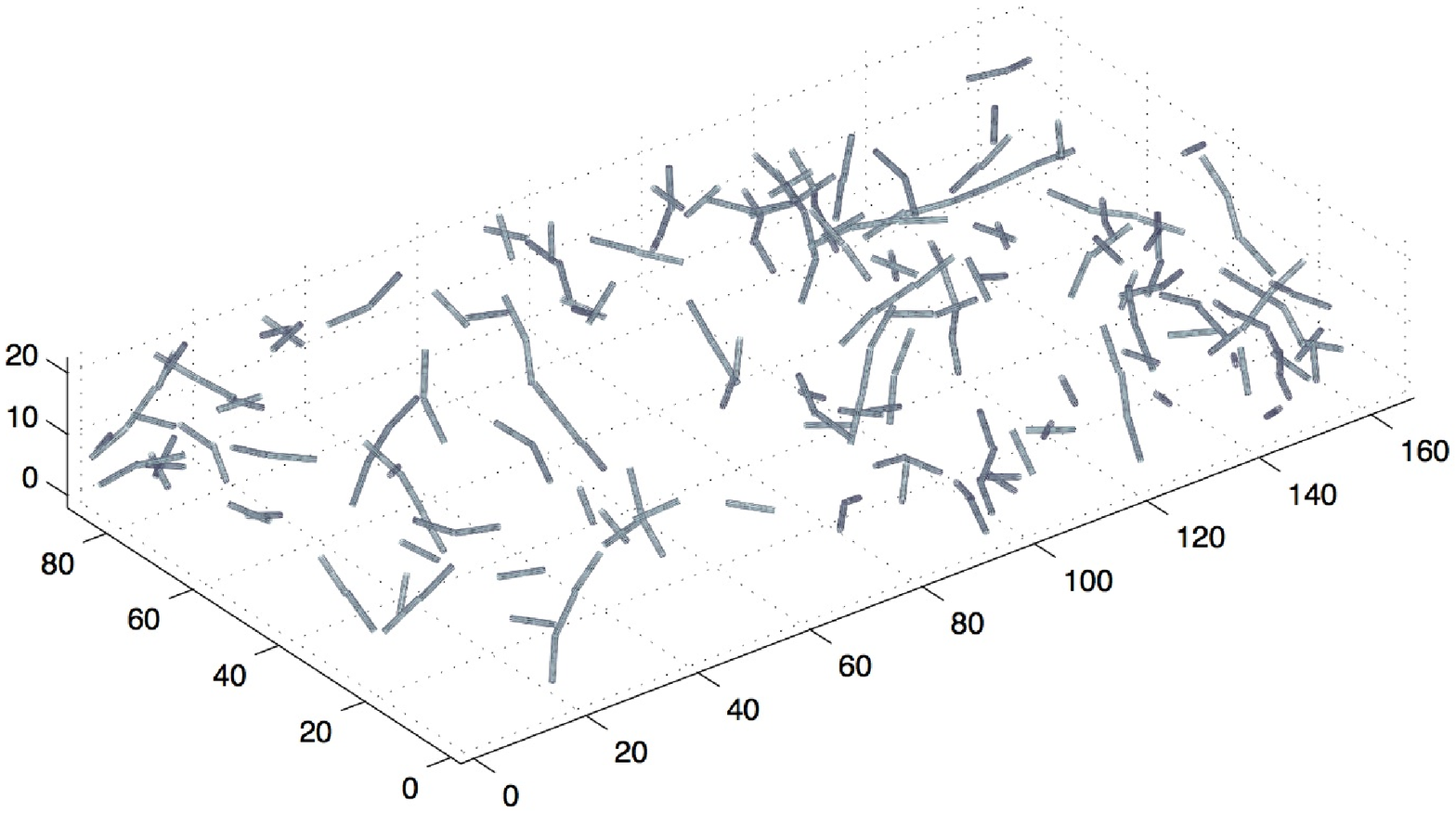}}\\ 
c)\resizebox{0.6\textwidth}{!}{\includegraphics*{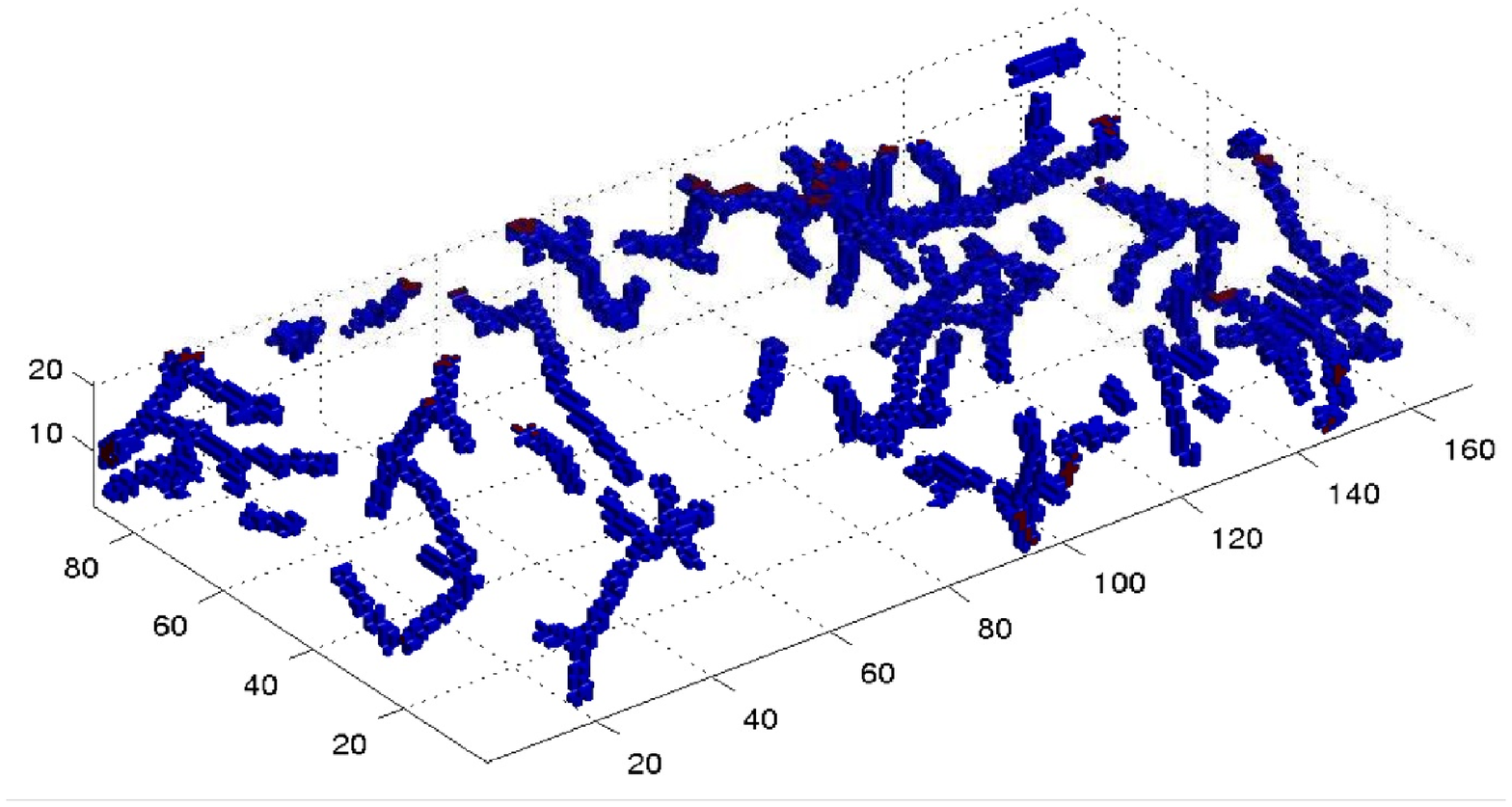}}\\ 
d)\resizebox{0.6\textwidth}{!}{\includegraphics*{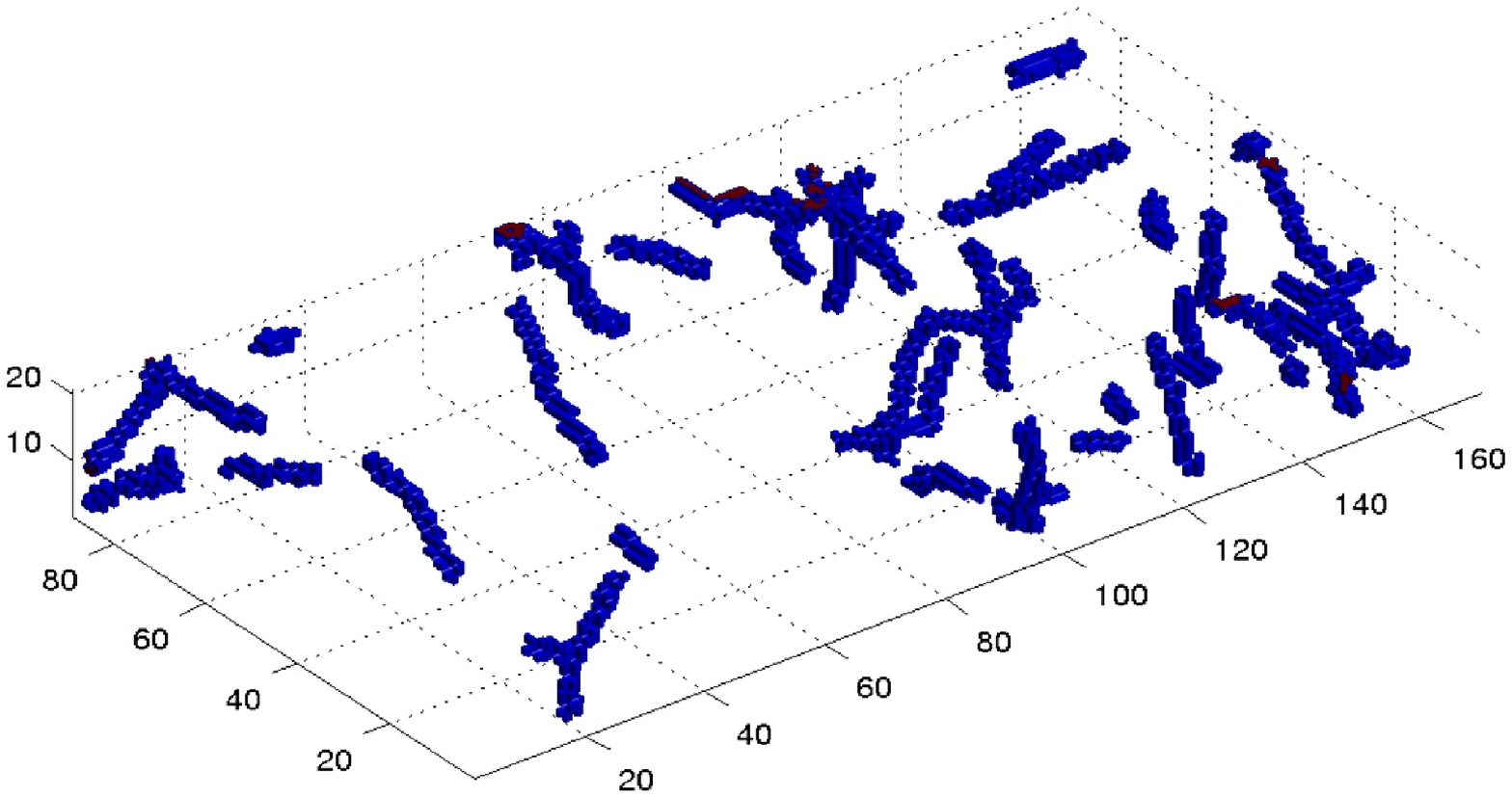}}\\ 
\caption{Data set $NGP250$. a) The data, 
b) Cylinder configuration obtained using the 
simulated annealing algorithm. Cover probabilities thresholded at c) 
$50\%$, % c) $75\%$, 
d) $95\%$.} 
\label{res_n250} 
\end{figure}

We ran the simulated annealing algorithm for $250000$ iterations. 
Samples were picked up every $250$ iterations. The obtained cylinder 
configurations for the data sets $NGP150$, $NGP200$ and $NGP250$ are shown 
in Figure~\ref{res_n150}b, Figure~\ref{res_n200}b and 
Figure~\ref{res_n250}b, respectively. 
 
The cylinders obtained after running the simulated annealing
are situated in the regions where  
we see that the data exhibit filamentary 
structures. Still, as simulated annealing requires infinitely many 
iterations till convergence, and also because of the fact that an 
infinity of solutions are proposed, we shall use coverage 
probabilities to ``average" the shape of the filaments. 
 
For each data set the corresponding domain $K$ was divided into 
small cubic cells of size $1\,h^{-1}\mbox{Mpc}$. 
These cells are small enough to be covered by a single cylinder. 
Computing the coverage probabilities directly using simulating 
annealing outputs is not possible since~\eqref{mcmc_cover_probability} 
is an expectation under $p(\yy,\theta)$. Hence, the algorithm was 
run until $T_0$, a fixed temperature close to zero, was reached. 
The coverage probabilities were computed as an expectation similar 
to~\eqref{mcmc_cover_probability} but under the law given 
by $p(\yy,\theta)^{1/T_{0}}$. The obtained result was 
thresholded using three distinct values: $0.5$, $0.75$ and $0.95$. 
For each of these values we obtain a map of the cells that have been
visited by our model with a frequency higher or equal than the given 
value. These visit maps for to the data sets $NGP150$, 
$NGP200$ and $NGP250$ are shown in Figure~\ref{res_n150}c-d, 
Figure~\ref{res_n200}c-d and Figure~\ref{res_n250}c-d respectively. 
 
The filamentary network is clearly outlined. Here we consider a 
filament the geometric structure obtained recursively from a visited 
cell and its corresponding neighbouring cells. In this case, we have 
defined the neighbours of a cell, the cells that have coordinates 
incremented or decremented by one with respect to the reference cell. 
Hence, a cell has $26$ neighbouring cells.

\begin{figure}[!htbp] 
\centering
\resizebox{0.6\textwidth}{!}{\includegraphics*{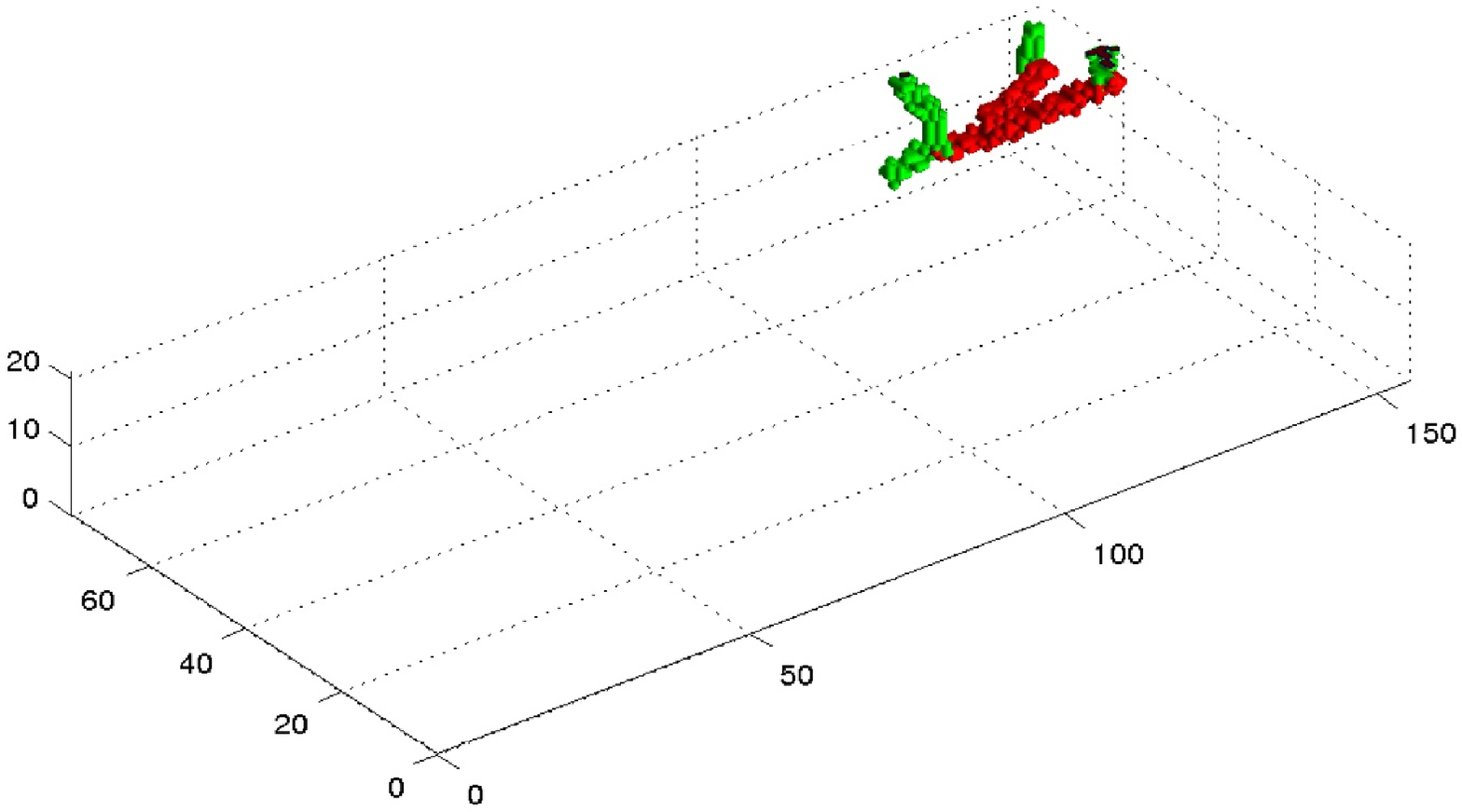}}\\ 
\caption{A simple filament from the $NGP250$ sample. 
Lighter shading (green colour) 
shows the 50\% cover probability threshold, darker shading (red colour) -- 
the 95\% threshold. 
\label{fil6_n250}} 
\end{figure}

\begin{figure}[!htbp] 
\centering
\resizebox{0.6\textwidth}{!}{\includegraphics*{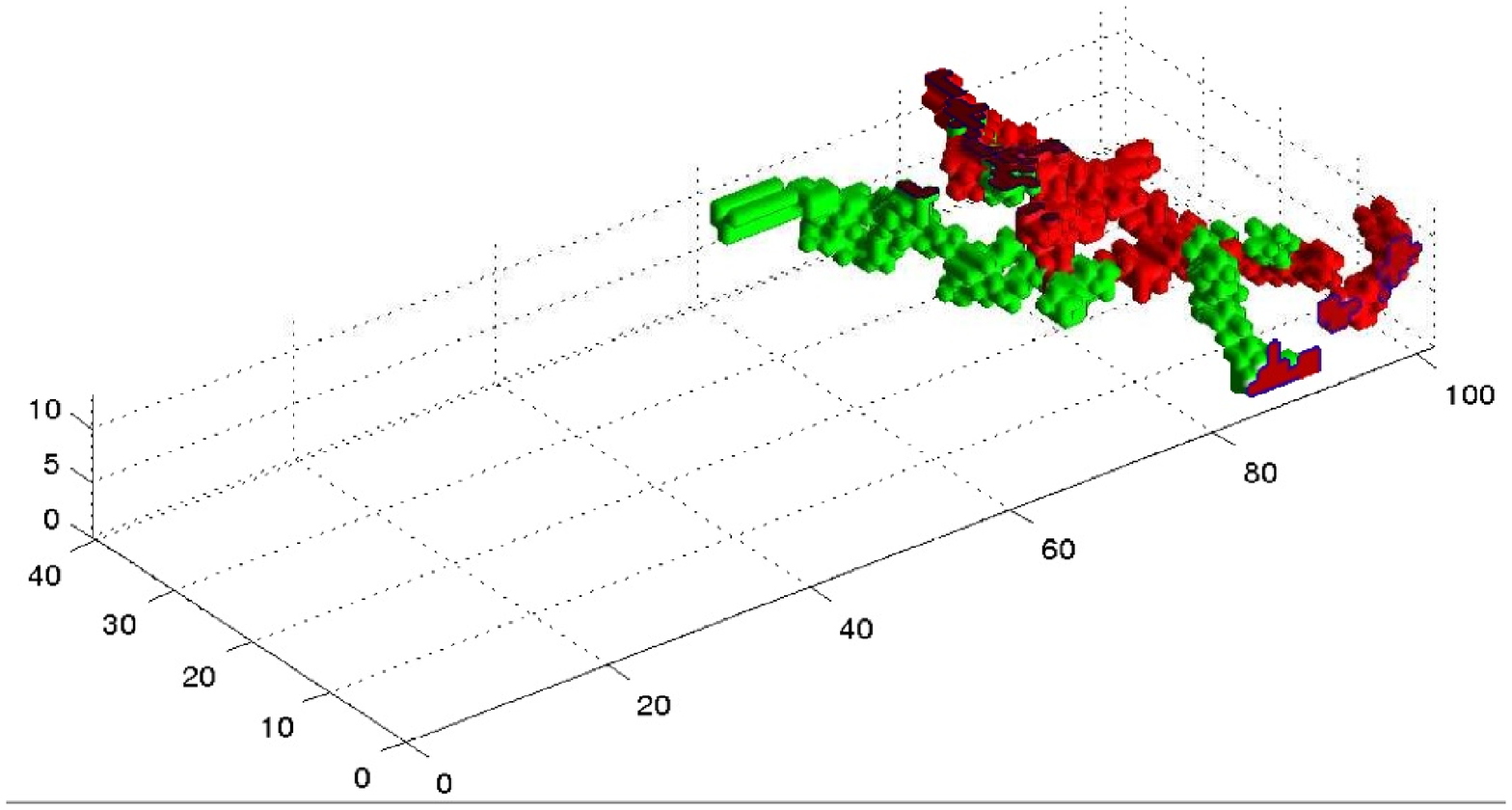}}\\ 
\caption{A complex filament from the $NGP150$ sample. 
Lighter shading (green colour) 
shows the 50\% cover probability threshold, darker shading (red colour) -- 
the 95\% threshold. 
\label{fil1_n150}} 
\end{figure}

Although many filaments are of simple form, there are many 
filaments which exhibit complex morphology, from simple branching 
to multibranch complexes. This is a new fact that will probably 
force cosmologists to reconsider their usual notion of filaments 
as simple bridges between clusters of galaxies. We show two examples 
of such filaments below, one of a comparatively simple shape 
(Fig.~\ref{fil6_n250}), and another of very complex shape 
(Fig.~\ref{fil1_n150}). 
 
\subsection{Statistics} 
The coverage probabilities are not a global test for our method, because 
these probabilities are computed only for small regions. The computation 
of the probability that a whole filament is covered by our model requires 
the knowledge a priori of the shape of the filament. (\cite{StoiGregMate05}) 
used for a different problem -- cluster detection in epidemiological data -- 
a test, to check that the pattern is detected rather because of the data 
than of the model parameters. 
 
Following this idea, the following experiments were carried out for each 
data set. First, the method was launched during $50000$ iterations at 
fixed $T=1.0$. Samples were picked up every $250$ iterations. The means of 
the sufficient statistics of the model were computed using these samples. The obtained 
results are shown in Table~\ref{tab:statdata}. 
 
\begin{table} 
\caption{The mean of the sufficient statistics for the three data sets:  
$\bar{n_2}$ is the mean number of the $2$-connected cylinders,  
$\bar{n_1}$ is the mean number of the $1$-connected cylinders and  
$\bar{n_0}$ is the mean number of the $0$-connected cylinders. \label{tab:statdata}}\\ 
\begin{tabular}{|l|r|r|r|} 
\hline 
& \multicolumn{3}{c|}{\strut Data sets}\\ 
\cline{2-4} \raisebox{1ex}[0pt]{Sufficient statistics}&\strut NGP150 & NGP200 
& NGP250\\ \hline 
\strut $\bar{n_2}$ & 4.13  & 5.83 & 9.88\\ 
$\bar{n_0}$ & 15.88 & 21.19 & 35.82\\ 
$\bar{n_1}$ & 21.35 & 35.58 & 46.49\\ 
\hline 
\end{tabular} 
\end{table} 
 
\begin{table} 
\caption{The maximum of mean of the sufficient statistics over binomial 
fields generated for each of the three data sets:  $\max\bar{n_2}$ is the maximum mean number of the $2$-connected cylinders,  $\max\bar{n_1}$ is the maximum mean number of the $1$-connected cylinders and  $\max\bar{n_0}$ is the maximum mean number of the $0$-connected cylinders. 
\label{tab:statbinomial}}\\ 
\begin{tabular}{|l|r|r|r|} 
\hline 
& \multicolumn{3}{c|}{\strut Binomial data sets}\\ 
\cline{2-4} \raisebox{1ex}[0pt]{Sufficient statistics}&\strut NGP150 & NGP200 
& NGP250\\ \hline 
\strut $\max \bar{n_2}$ & 0.015  & 0.05 & 0.015\\ 
$\max \bar{n_0}$ & 0.54 & 0.27 & 0.45\\ 
$\max \bar{n_1}$ & 0.39 & 0.24 & 0.33\\ 
\hline 
\end{tabular} 
\end{table} 
 
The second experiment consisted of re-distributing uniformly the 
points inside the domain $K$. So, the points follow a binomial 
distribution. For each data set this operation was done $100$ times, 
hence obtaining $100$ point fields. For each point field the method 
was launched in the conditions previously described. The mean of the 
sufficient statistics was then computed. The maximum values over all 
the $100$ means for each data set are shown in 
Table~\ref{tab:statbinomial}. 
 
These results clearly indicate that the original data exhibit a 
filamentary structure. No filamentary structure is detected when the same 
model runs over binomial fields of points that has the same number of 
points as the original data. This test discriminates the two cases 
at a $p$-value less than 1\%, and assures us that the filaments we 
find are due to data and not to our model. 
 
\section{Conclusion and perspectives}

We have applied an object point process -- Bisous model -- to 
objectively find filaments in galaxy redshift surveys 
(three-dimensional galaxy maps). For that, we defined the model, 
fixed some of the interaction parameters and chose priors for the 
remaining parameters. The definition of the data term is very 
intuitive and rather a simple test. Much more elaborated methods 
testing the alignment of the points along the direction of the 
cylinder against the completely spatial randomness need 
investigation. The uniform law for the interaction parameters was 
preferred in order to give the same chances to a wide range of 
topologies of the filamentary network. Still, if concrete prior 
information about the topology of the filamentary network is 
available then this should be integrated in the model. 
 
We have run simulated annealing sequences and select filaments on 
the basis of coverage probabilities for individual cells of sample 
volume. The coverage probabilities are to be seen as a way of 
averaging the shape of the filamentary structure. They have the 
advantage to allow inference from statistics instead of a single 
realisation. Their main drawback is that the coverage probabilities 
are computed locally, for small regions. Still, the visualisation of 
the visit maps built on these probabilities brings new ideas and 
hypotheses about the topologies of the cosmic filaments. A global 
Monte Carlo statistical test was built to test the existence of the 
filaments in the data. To do this, we have calculated sufficient 
statistics for the data sets and made a comparison with the 
sufficient statistics obtained on binomial point fields having the 
same number of points as the data. Our test was indicating that the 
filaments we find are defined by the data, not by the chosen model. 
 
The method used in this paper can be extended in different ways. One 
natural extension is to use different generating elements instead on 
cylinders, e.g., planar elements or clusters. 
(\cite{StoiGregMate05}). Although traces of planar structures are 
seen in superclusters of galaxies, these have been difficult to 
quantify, mainly because of their low density contrast. Another 
interesting application is to search for dynamically bound groups 
and clusters of galaxies that have a typical 'finger-of-God' 
signature in redshift space, extended along the line-of-sight. And 
there remains a question if the method could be extended to 
inhomogeneous point processes -- this would allow us to use all the 
observational data, not only volume-limited subsamples. 
 
\section{Acknowledgements}

We thank our referees for their thorough analysis and constructive comments.
This work has been supported by the University of Valencia through a 
visiting professorship for Enn Saar, by the Spanish MCyT project 
AYA2003-08739-C02-01 (including FEDER) and AYA2006-14056, by 
the Estonian Ministry of Education and Science, research project 
TO 0062465s03, and by the Estonian Science Foundation grant 6104. 
 
%\bibliography{astro3d} 

\begin{thebibliography}{}

\bibitem[\protect\citeauthoryear{Barrow, Sonoda, and Bhavsar}{Barrow
  et~al.}{1985}]{barrow85}
Barrow, J.~D., D.~H. Sonoda, and S.~P. Bhavsar (1985).
\newblock Minimal spanning tree, filaments and galaxy clustering.
\newblock {\em \mnras\/}~{\em 216}, 17--35.

\bibitem[\protect\citeauthoryear{Berthelsen and M{\o}ller}{Berthelsen and
  M{\o}ller}{2002}]{BertMoll02}
Berthelsen, K.~K. and J.~M{\o}ller (2002).
\newblock A primer on perfect simulation for spatial point processes.
\newblock {\em Bulletin of the Brazilian Mathematical Society\/}~{\em 33},
  351--367.

\bibitem[\protect\citeauthoryear{Bharadwaj, Bhavsar, and Sheth}{Bharadwaj
  et~al.}{2004}]{bhavsar03}
Bharadwaj, S., S.~P. Bhavsar, and J.~V. Sheth (2004).
\newblock The size of the longest filaments in the universe.
\newblock {\em \apj\/}~{\em 606}, 25--31.

\bibitem[\protect\citeauthoryear{{Bharadwaj} and {Pandey}}{{Bharadwaj} and
  {Pandey}}{2004}]{bharadwaj04}
{Bharadwaj}, S. and B.~{Pandey} (2004, November).
\newblock {Using the Filaments in the Las Campanas Redshift Survey to Test the
  {$\Lambda$}CDM Model}.
\newblock {\em \apj\/}~{\em 615}, 1--6.

\bibitem[\protect\citeauthoryear{{Colberg}, {Krughoff}, and
  {Connolly}}{{Colberg} et~al.}{2005}]{colberg05}
{Colberg}, J.~M., K.~S. {Krughoff}, and A.~J. {Connolly} (2005, May).
\newblock {Intercluster filaments in a {$\Lambda$}CDM Universe}.
\newblock {\em \mnras\/}~{\em 359}, 272--282.

\bibitem[\protect\citeauthoryear{{Colless}, {Dalton}, {Maddox}, {Sutherland},
  {Norberg}, {Cole}, {Bland-Hawthorn}, {Bridges}, {Cannon}, {Collins}, {Couch},
  {Cross}, {Deeley}, {De Propris}, {Driver}, {Efstathiou}, {Ellis}, {Frenk},
  {Glazebrook}, {Jackson}, {Lahav}, {Lewis}, {Lumsden}, {Madgwick}, {Peacock},
  {Peterson}, {Price}, {Seaborne}, and {Taylor}}{{Colless}
  et~al.}{2001}]{2dFGRS}
{Colless}, M., G.~{Dalton}, S.~{Maddox}, W.~{Sutherland}, P.~{Norberg},
  S.~{Cole}, J.~{Bland-Hawthorn}, T.~{Bridges}, R.~{Cannon}, C.~{Collins},
  W.~{Couch}, N.~{Cross}, K.~{Deeley}, R.~{De Propris}, S.~P. {Driver},
  G.~{Efstathiou}, R.~S. {Ellis}, C.~S. {Frenk}, K.~{Glazebrook}, C.~{Jackson},
  O.~{Lahav}, I.~{Lewis}, S.~{Lumsden}, D.~{Madgwick}, J.~A. {Peacock}, B.~A.
  {Peterson}, I.~{Price}, M.~{Seaborne}, and K.~{Taylor} (2001, December).
\newblock {The 2dF Galaxy Redshift Survey: spectra and redshifts}.
\newblock {\em \mnras\/}~{\em 328}, 1039--1063.

\bibitem[\protect\citeauthoryear{{Croton}, {Colless}, {Gazta{\~n}aga}, {Baugh},
  {Norberg}, {Baldry}, {Bland-Hawthorn}, {Bridges}, {Cannon}, {Cole},
  {Collins}, {Couch}, {Dalton}, {de Propris}, {Driver}, {Efstathiou}, {Ellis},
  {Frenk}, {Glazebrook}, {Jackson}, {Lahav}, {Lewis}, {Lumsden}, {Maddox},
  {Madgwick}, {Peacock}, {Peterson}, {Sutherland}, and {Taylor}}{{Croton}
  et~al.}{2004}]{croton1}
{Croton}, D.~J., M.~{Colless}, E.~{Gazta{\~n}aga}, C.~M. {Baugh}, P.~{Norberg},
  I.~K. {Baldry}, J.~{Bland-Hawthorn}, T.~{Bridges}, R.~{Cannon}, S.~{Cole},
  C.~{Collins}, W.~{Couch}, G.~{Dalton}, R.~{de Propris}, S.~P. {Driver},
  G.~{Efstathiou}, R.~S. {Ellis}, C.~S. {Frenk}, K.~{Glazebrook}, C.~{Jackson},
  O.~{Lahav}, I.~{Lewis}, S.~{Lumsden}, S.~{Maddox}, D.~{Madgwick}, J.~A.
  {Peacock}, B.~A. {Peterson}, W.~{Sutherland}, and K.~{Taylor} (2004, August).
\newblock {The 2dF Galaxy Redshift Survey: voids and hierarchical scaling
  models}.
\newblock {\em \mnras\/}~{\em 352}, 828--836.

\bibitem[\protect\citeauthoryear{{Eriksen}, {Novikov}, {Lilje}, {Banday}, and
  {G{\' o}rski}}{{Eriksen} et~al.}{2004}]{eriksen04}
{Eriksen}, H.~K., D.~I. {Novikov}, P.~B. {Lilje}, A.~J. {Banday}, and K.~M.
  {G{\' o}rski} (2004, September).
\newblock {Testing for Non-Gaussianity in the Wilkinson Microwave Anisotropy
  Probe Data: Minkowski Functionals and the Length of the Skeleton}.
\newblock {\em \apj\/}~{\em 612}, 64--80.

\bibitem[\protect\citeauthoryear{Geyer}{Geyer}{1994}]{Geye94}
Geyer, C.~J. (1994).
\newblock On the convergence of monte carlo maximum likelihood calculations.
\newblock {\em Journal of the Royal Statistical Society, Series B\/}~{\em 56},
  261--274.

\bibitem[\protect\citeauthoryear{Geyer}{Geyer}{1999}]{Geye99}
Geyer, C.~J. (1999).
\newblock Likelihood inference for spatial point processes.
\newblock In O.~Barndorff-Nielsen, W.~S. Kendall, and M.~N.~M. van Lieshout
  (Eds.), {\em Stochastic geometry, likelihood and computation}. CRC
  Press/Chapman and Hall, Boca Raton.

\bibitem[\protect\citeauthoryear{Geyer and M{\o}ller}{Geyer and
  M{\o}ller}{1994}]{GeyeMoll94}
Geyer, C.~J. and J.~M{\o}ller (1994).
\newblock Simulation procedures and likelihood inference for spatial point
  processes.
\newblock {\em Scan. J. Stat.\/}~{\em 21}, 359--373.

\bibitem[\protect\citeauthoryear{Geyer and Thompson}{Geyer and
  Thompson}{1992}]{GeyeThom92}
Geyer, C.~J. and E.~A. Thompson (1992).
\newblock Constrained monte carlo maximum likelihood for dependent data.
\newblock {\em Journal of the Royal Statistical Society, Series B\/}~{\em 54},
  657--699.

\bibitem[\protect\citeauthoryear{Green}{Green}{1995}]{Gree95}
Green, P. (1995).
\newblock Reversible jump {MCMC} computation and bayesian model determination.
\newblock {\em Biometrika\/}~{\em 82}, 711--732.

\bibitem[\protect\citeauthoryear{Hearn and Baker}{Hearn and
  Baker}{1994}]{HearBake94}
Hearn, D. and M.~P. Baker (1994).
\newblock {\em Computer graphics. Second edition}.
\newblock Prentice-Hall, Englewood Cilffs, NJ.

\bibitem[\protect\citeauthoryear{Kendall and M{\o}ller}{Kendall and
  M{\o}ller}{2000}]{KendMoll00}
Kendall, W.~S. and J.~M{\o}ller (2000).
\newblock Perfect simulation using dominating processes on ordered spaces, with
  application to locally stable point processes.
\newblock {\em Adv. Appl. Prob.\/}~{\em 32}, 844--865.

\bibitem[\protect\citeauthoryear{Lacoste, Descombes, and Zerubia}{Lacoste
  et~al.}{2005}]{LacoDescZeru05}
Lacoste, C., X.~Descombes, and J.~Zerubia (2005).
\newblock Point processes for unsupervised line network extraction in remote
  sensing.
\newblock {\em IEEE Trans. Pattern Analysis and Machine Intelligence,\/}~{\em
  27}, 1568--1579.

\bibitem[\protect\citeauthoryear{Mart{\'\i}nez and Saar}{Mart{\'\i}nez and
  Saar}{2002}]{martsaar02}
Mart{\'\i}nez, V.~J. and E.~Saar (2002).
\newblock {\em Statistics of the Galaxy Distribution}.
\newblock Chapman {\&} Hall/CRC, Boca Raton.

\bibitem[\protect\citeauthoryear{M{\o}ller, Pettitt, Berthelsen, and
  Reeves}{M{\o}ller et~al.}{2004}]{MollPettBertReev04}
M{\o}ller, J., A.~N. Pettitt, K.~K. Berthelsen, and R.~W. Reeves (2004).
\newblock An efficient markov chain monte carlo method for distributions with
  intractable normalizing constants.
\newblock Research report R-2004-02, Department of Mathematical Sciences,
  Aalborg University.

\bibitem[\protect\citeauthoryear{M{\o}ller and Waagepetersen}{M{\o}ller and
  Waagepetersen}{2003}]{MollWaag03}
M{\o}ller, J. and R.~P. Waagepetersen (2003).
\newblock {\em Statistical inference for spatial point processes}.
\newblock Chapman {\&} Hall/CRC, Boca Raton.

\bibitem[\protect\citeauthoryear{{Novikov}, {Colombi}, and
  {Dor{\'e}}}{{Novikov} et~al.}{2006}]{novikov06}
{Novikov}, D., S.~{Colombi}, and O.~{Dor{\'e}} (2006).
\newblock {Skeleton as a probe of the cosmic web: the two-dimensional case}.
\newblock {\em \mnras\/}~{\em 366}, 1201--1216.

\bibitem[\protect\citeauthoryear{{Pandey} and {Bharadwaj}}{{Pandey} and
  {Bharadwaj}}{2005}]{pandey05}
{Pandey}, B. and S.~{Bharadwaj} (2005, March).
\newblock {A two-dimensional analysis of percolation and filamentarity in the
  Sloan Digital Sky Survey Data Release One}.
\newblock {\em \mnras\/}~{\em 357}, 1068--1076.

\bibitem[\protect\citeauthoryear{{Pimbblet}}{{Pimbblet}}{2005}]{pimbblet05}
{Pimbblet}, K.~A. (2005).
\newblock {Pulling Out Threads from the Cosmic Tapestry: Defining Filaments of
  Galaxies}.
\newblock {\em Publications of the Astronomical Society of Australia\/}~{\em
  22}, 136--143.

\bibitem[\protect\citeauthoryear{{Pimbblet} and {Drinkwater}}{{Pimbblet} and
  {Drinkwater}}{2004}]{pimbblet04a}
{Pimbblet}, K.~A. and M.~J. {Drinkwater} (2004, January).
\newblock {Intercluster Filaments of Galaxies Programme: pilot study survey and
  results}.
\newblock {\em \mnras\/}~{\em 347}, 137--143.

\bibitem[\protect\citeauthoryear{{Pimbblet}, {Drinkwater}, and
  {Hawkrigg}}{{Pimbblet} et~al.}{2004}]{pimbblet04b}
{Pimbblet}, K.~A., M.~J. {Drinkwater}, and M.~C. {Hawkrigg} (2004, November).
\newblock {Intercluster filaments of galaxies programme: abundance and
  distribution of filaments in the 2dFGRS catalogue}.
\newblock {\em \mnras\/}~{\em 354}, L61.

\bibitem[\protect\citeauthoryear{Preston}{Preston}{1977}]{Pres77}
Preston, C.~J. (1977).
\newblock Spatial birth-and-death processes.
\newblock {\em Bull. Int. Stat. Inst.\/}~{\em 46}, 371--391.

\bibitem[\protect\citeauthoryear{Reiss}{Reiss}{1993}]{Reis93}
Reiss, R.~D. (1993).
\newblock {\em A course on point processes}.
\newblock Springer-Verlag, New York.

\bibitem[\protect\citeauthoryear{{Sousbie}, {Pichon}, {Courtois}, {Colombi},
  and {Novikov}}{{Sousbie} et~al.}{2006}]{sousbie06}
{Sousbie}, T., C.~{Pichon}, H.~{Courtois}, S.~{Colombi}, and D.~{Novikov}
  (2006).
\newblock {The 3D skeleton of the SDSS}.
\newblock {\em astro-ph/0602628\/}.

\bibitem[\protect\citeauthoryear{Stoica, Descombes, van Lieshout, and
  Zerubia}{Stoica et~al.}{2002}]{StoiDescLiesZeru02}
Stoica, R.~S., X.~Descombes, M.~N.~M. van Lieshout, and J.~Zerubia (2002).
\newblock An application of marked point processes to the extraction of linear
  networks from images.
\newblock In J.~Mateu and F.~Montes (Eds.), {\em Spatial statistics through
  applications}. WIT Press, Southampton, UK.

\bibitem[\protect\citeauthoryear{Stoica, Descombes, and Zerubia}{Stoica
  et~al.}{2004}]{StoiDescZeru04}
Stoica, R.~S., X.~Descombes, and J.~Zerubia (2004).
\newblock A {Gibbs} point process for road extraction in remotely sensed
  images.
\newblock {\em Int. J. Computer Vision\/}~{\em 57(2)}, 121--136.

\bibitem[\protect\citeauthoryear{Stoica and Gay}{Stoica and
  Gay}{2005}]{StoiGay05}
Stoica, R.~S. and E.~Gay (2005, July).
\newblock Cluster detection in spatial data using marked point processes.
\newblock Research report 11-2005, INRA, Avignon.

\bibitem[\protect\citeauthoryear{Stoica, Gregori, and Mateu}{Stoica
  et~al.}{2005a}]{StoiGregMate05}
Stoica, R.~S., P.~Gregori, and J.~Mateu (2005a).
\newblock Simulated annealing and object point processes : tools for analysis
  of spatial patterns.
\newblock {\em Stochastic Processes and their Applications\/}~{\em 115},
  1860--1882.

\bibitem[\protect\citeauthoryear{Stoica, Martinez, Mateu, and Saar}{Stoica
  et~al.}{2005b}]{StoiMartMateSaar05}
Stoica, R.~S., V.~J. Martinez, J.~Mateu, and E.~Saar (2005b).
\newblock Detection of cosmic filaments using the Candy model.
\newblock {\em Astronomy and Astrophysics\/}~{\em 434}, 423--432.

\bibitem[\protect\citeauthoryear{van Lieshout}{van Lieshout}{1994}]{Lies94}
van Lieshout, M. N.~M. (1994).
\newblock Stochastic annealing for nearest-neighbour point processes with
  application to object recognition.
\newblock {\em Adv. Appl. Prob.\/}~{\em 26}, 281--300.

\bibitem[\protect\citeauthoryear{van Lieshout}{van Lieshout}{2000}]{Lies00}
van Lieshout, M. N.~M. (2000).
\newblock {\em Markov point processes and their applications}.
\newblock Imperial College Press/World Scientific Publishing, London/Singapore.

\bibitem[\protect\citeauthoryear{van Lieshout and Stoica}{van Lieshout and
  Stoica}{2003a}]{LiesStoi03}
van Lieshout, M. N.~M. and R.~S. Stoica (2003a).
\newblock The {Candy} model revisited: properties and inference.
\newblock {\em Stat. Neerlandica\/}~{\em 57}, 1--30.

\bibitem[\protect\citeauthoryear{van Lieshout and Stoica}{van Lieshout and
  Stoica}{2003b}]{LiesStoi03a}
van Lieshout, M. N.~M. and R.~S. Stoica (2003b).
\newblock Perfect simulation for marked point processes.
\newblock Research report PNA-0306, CWI.

\bibitem[\protect\citeauthoryear{van Lieshout and Stoica}{van Lieshout and
  Stoica}{2006}]{LiesStoi06}
van Lieshout, M. N.~M. and R.~S. Stoica (2006).
\newblock Perfect simulation for marked point processes.
\newblock {\em Computational Statistics and Data Analysis\/}~{\em 51}, 679--698. 

\end{thebibliography}

\end{document}